\RequirePackage{fix-cm}
\documentclass[twocolumn,epjc3]{svjour3}  
\smartqed  
\RequirePackage{graphicx}
\usepackage{soul}
\usepackage{amssymb}
\RequirePackage{booktabs}
\usepackage{amsmath}
\usepackage{xcolor}
\usepackage{lineno}
\usepackage{hyperref}
\usepackage{float}
\usepackage{cite}
\usepackage{tikz}

\definecolor{bleudefrance}{rgb}{0.19, 0.55, 0.91}

\nolinenumbers
\RequirePackage{amsmath}
\RequirePackage{verbatim}
\usepackage{hyperref}
\hypersetup{colorlinks,breaklinks,
            linkcolor=blue,urlcolor=blue,
            anchorcolor=blue,citecolor=blue}
            
\journalname{Eur. Phys. J. C}
\begin{document}

\title{Searches for new physics below twice the electron mass with GERDA 
}

\author{
The \mbox{\protect{\sc{Gerda}}} collaboration\thanksref{corrauthor}
\and  \\[4mm]
%
M.~Agostini\thanksref{UCL} \and
A.~Alexander\thanksref{UCL} \and
G.~Araujo\thanksref{UZH} \and
A.M.~Bakalyarov\thanksref{KU} \and
M.~Balata\thanksref{ALNGS} \and
I.~Barabanov\thanksref{INRM,deceased} \and
L.~Baudis\thanksref{UZH} \and
C.~Bauer\thanksref{HD} \and
S.~Belogurov\thanksref{ITEP,INRM,alsoMEPHI} \and
A.~Bettini\thanksref{PDUNI,PDINFN}
% ...
}
\authorrunning{the \textsc{Gerda} collaboration}
\thankstext{corrauthor}{
  \emph{correspondence:}  gerda-eb@mpi-hd.mpg.de}
\thankstext{deceased}{\emph{deceased}}
\thankstext{alsoMEPHI}{\emph{also at:} NRNU MEPhI, Moscow, Russia}
\thankstext{nowDuke}{\emph{present address:} Duke University, Durham, NC USA}
\thankstext{alsoLev}{\emph{also at:} Moscow Inst. of Physics and Technology,
  Russia}
\thankstext{nowIKZ}{\emph{present address:} Semilab Zrt, Budapest, Hungary}
\thankstext{alsoDubna}{\emph{also at:} Dubna State University, Dubna, Russia}
\thankstext{nowBerkeley}{\emph{present address:} Nuclear Science Division, Berkeley, USA}
\institute{ 
INFN Laboratori Nazionali del Gran Sasso, Assergi, Italy\label{ALNGS}
% ...
\and
Max-Planck-Institut f{\"u}r Kernphysik, Heidelberg, Germany\label{HD} \and
Department of Physics and Astronomy, University College London, London, UK\label{UCL}
% ...
\and
Institute for Nuclear Research of the Russian Academy of Sciences, Moscow, Russia\label{INRM} \and
Institute for Theoretical and Experimental Physics, NRC ``Kurchatov Institute'', Moscow, Russia\label{ITEP} \and
National Research Centre ``Kurchatov Institute'', Moscow, Russia\label{KU}
% ...
\and
Dipartimento di Fisica e Astronomia, Universit{\`a} degli Studi di 
Padova, Padua, Italy\label{PDUNI} \and
INFN  Padova, Padua, Italy\label{PDINFN}
% ...
\and
Physik-Institut, Universit{\"a}t Z{\"u}rich, Z{u}rich, Switzerland\label{UZH}
} 
%


\date{ Received: date / Accepted: date}

\maketitle

\begin{abstract}
A search for full energy depositions from bosonic keV-scale dark matter candidates
of masses between 65~keV and 1021~keV has been performed with data collected during Phase~II of the 
GERmanium Detector Array (\textsc{Gerda}) experiment. 
Our analysis includes direct dark matter absorption as well as  dark Compton scattering. 
With a total exposure of 105.5~kg\,yr, no evidence for a signal above the background has been observed. 
The resulting exclusion limits deduced with either Bayesian or Frequentist statistics are the  most stringent direct constraints in the major part of the 140-1021~keV mass range. 
As an example, at a mass of 150~keV the dimensionless coupling of dark photons and axion-like particles to electrons has been constrained to $\alpha ' / \alpha<8.7~\times 10^{-24}$ and $g_{\rm ae}<3.3~\times 10^{-12}$ at 90\% credible interval (CI), respectively. \\
\noindent Additionally, a search for peak-like signals from beyond the Standard Model decays of nucleons and electrons is performed.
We find for the inclusive decay of a single neutron in $^{76}$Ge a lower lifetime limit of $\tau_{\rm n} > 1.5 \times 10^{24}$ yr and for a proton $\tau_{\rm p} > 1.3 \times 10^{24}$ yr at 90\% CI. For the electron decay $e^\text{-} \rightarrow \nu_{\rm e} \gamma$ a lower limit of $\tau_{\rm e} > 5.4\times 10^{25}$ yr at 90\% CI has been determined.
\end{abstract}


\section{Introduction}
\label{sec:intro}
\indent The main goal of the \textsc{Gerda} experiment was to search for the neutrinoless double-beta ($0\nu\beta\beta$) decay of $^{76}$Ge. An array of high-purity germanium (HPGe) detectors enriched up to $\sim$87\% in $^{76}$Ge was employed in an active liquid argon (LAr) shield. The shielded environment and the excellent energy resolution of the  Ge detectors made the experiment also suitable for the search of peak-like signatures induced by new physics processes other than $0\nu\beta\beta$ decay. In this paper, searches for keV-scale bosonic dark matter (DM) interactions and single-particle disappearance processes are reported.\\
\indent \textsc{Gerda} is sensitive to pseudoscalar (axion-like particles, ALPs) and vector (dark photons, DPs) bosonic DM candidates, sometimes referred to as super Weakly Interacting Massive Particles (superWIMPs)~\cite{Pospelov:2008jk}. A previous search for photoelectric-like absorption of bosonic DM candidates, with masses\footnote{In this paper, natural units are used, i.e. $c = 1$.} up to 1~MeV, was reported by \textsc{Gerda} in \cite{GERDA:2020emj}. In this paper, a second interaction process, i.e. the dark Compton scattering process, was included in the calculation of the interaction rate of these DM particles with electrons~\cite{PhysRevD.104.083030,PhysRevLett.128.191801}. Despite its lower detection efficiency at higher masses (see Table~\ref{tab:effi}), the dark Compton scattering benefits from a larger interaction cross-section for energies above $\sim$140~keV~\cite{PhysRevD.104.083030}.\\
\indent Moreover, the experiment can probe beyond the Standard Model (BSM) decay processes violating conservation laws of the Standard Model (SM), e.g., the decay of a single neutron or proton~\cite{PhysRevD.101.015005}. As pointed out by Sakharov, the violation of the conservation of baryon number is one of the three fundamental criteria needed to be fulfilled to produce the matter-antimatter asymmetry in the early Universe~\cite{Sakharov:1967dj}. 
\textsc{Gerda} explores the disappearance of a single nucleon in $^{76}$Ge by looking for the $\beta$-decay of the $^{75}$Ge ground state to an excited state of $^{75}$As in coincidence with the $\gamma$-ray emitted in the subsequent $^{75}$As de-excitation.
The population of the $^{75}$Ge ground state follows the disappearance of either a neutron or a proton in $^{76}$Ge. Proton decay, in particular, populates first the unstable $^{75}$Ga nucleus that later decays by $\beta$-emission to $^{75}$Ge.\\   
\indent Another BSM process of interest is the decay of an electron via $e^\text{-}\rightarrow\nu_{\rm e}\nu_{\rm e}\nu_{\rm e}$ or $e^\text{-} \rightarrow \nu_{\rm e} \gamma$, where the latter channel is explored in this study. It allows a sensitive test of the U(1) gauge symmetry that ensures the stability of the electron as well as the zero mass of the photon.\\
\noindent The paper is structured as follows. In Sect.~\ref{sec:Physics searches}, the theoretical framework for the bosonic DM and single-particle disappearance searches are introduced. In Sect.~\ref{sec:data_efficiency} an overview of the \textsc{Gerda} setup is given, focusing on the data selection and the evaluation of detection efficiencies for the final states of interest. In Sect.~\ref{sec:methods}, Frequentist and Bayesian analysis methods, are sketched that are used in our data analysis. In Sect.~\ref{sec:results}, results obtained with both statistical frameworks are presented. Conclusions are drawn in Sect.~\ref{sec:Conclusions}.

\section{Approaches to the search for new physics}
\label{sec:Physics searches}

\subsection{Bosonic dark matter}
\label{subsec:Bosonic dark matter}

Several galactic and cosmological observations indicate the existence of DM. However, its nature is still unknown. In the cosmological standard model $\Lambda$CDM the energy density contains 27\% of DM, with the rest being ordinary matter (5\%) and dark energy (68\%). 
Hence, several laboratory studies have been conducted or are planned to detect and investigate the nature of DM~\cite{Mitsou:2019xzu}. 
Various theoretical models for DM candidates have been proposed for masses ranging over many orders of magnitudes~\cite{EuropeanStrategyforParticlePhysicsPreparatoryGroup}. In the energy range explored by \textsc{Gerda}, bosonic keV-scale DM particles are particularly interesting candidates. Masses within this range imply a super-weak interaction strength between the DM and the SM sector, much weaker than normal weak-scale interactions. The mass and the cross-section requirements follow directly from the necessity of having an early thermal decoupling of the DM sector, which happened before the electroweak epoch at $T_{\rm EW}\sim100\,\mathrm{GeV}$~\cite{Pospelov:2008jk}. In this paper, pseudoscalar and vector bosonic DM candidates are considered, focusing on masses below $2m_{\rm e}\sim1022$ keV, where $m_{\rm e}$ is the electron mass. For DM masses $m_{\rm DM} \geq 2m_{\rm e}$, decays into $e^{\text{-}}e^{\text{+}}$ pairs are possible, making long-lived DM highly unlikely. Below this threshold, bosonic DM candidates are stable at the tree level.
In addition, radiative decays of ALPs and DPs into photons are possible at loop level in the keV-MeV range~\cite{Pospelov:2008jk,PhysRevLett.128.221302}.

\indent The previous \textsc{Gerda} study focused on the bosonic DM absorption in processes 
analogous to the photoelectric effect. Here, the DM particle is completely absorbed by a 
detector's atom, which later releases an electron in the final state. The expected signal is a 
full absorption peak at the rest mass of the DM, assuming these DM particles have very small 
kinetic energies at $\beta=v_{\rm DM}\sim10^{-3}$. The peak is then broadened due to the 
detector's energy resolution. 
The photoelectric-like absorption cross section at a given mass is \cite{Pospelov:2008jk}
\begin{equation}
    \sigma_{\rm a,e}\left(m_{\rm a} \right) = g_{\rm ae}^2\,
    \frac{m_{\rm a}^2\,\sigma_{\rm pe}(m_{\rm a})}
    {\beta}\left(\frac{3}{16\pi\alpha m_{\rm e}^2} \right)
\end{equation}
and 
\begin{equation}
    \sigma_{\rm V,e}\left(m_{\rm V} \right) = \frac{\alpha^{'}}{\alpha}\,
    \frac{\sigma_{\rm pe}(m_{\rm V})}
    {\beta}
\end{equation}
for pseudoscalar and vector DM candidates, respectively. Here, $m_{\rm a}$ ($m_{\rm V}$) is the 
ALP (DP) mass and $\sigma_{\rm pe}$ is the energy-dependent photoelectric cross-section of Ge. 
Assuming a DM density of $\rho_{\rm DM}=0.3$~GeV cm$^{-3}$ and a corresponding average DM flux 
$\Phi_{\rm DM}$ per barn (b) and day (d) at Earth \cite{Majorana:2016hop},
\begin{equation}
    \Phi_{\rm DM}\left( m_{\rm DM} \right) = \beta \, \frac{7.8\times10^{-4}}
    {m_{\rm DM}/{\rm [keV]}} \,\text{b}^{-1}\,\text{d}^{-1}\,,
\end{equation}
above cross sections are converted to
the absorption interaction rate for pseudoscalar and vector DM, respectively, \cite{GERDA:2020emj}
\begin{equation}
    R_{\rm a}^{\rm A}  = \frac{1.47\times10^{19}}{M_{\rm tot}}\,
    g_{\rm ae}^2
    \left(\frac{m_{\rm a}}{\left[\text{keV} \right]} \right)
    \left(\frac{\sigma_{\rm pe}}{\left[\text{b} \right]} \right)\,
    \text{kg}^{-1}\,\text{d}^{-1}
\label{eq:R_pseudo_abs}
\end{equation}
and
\begin{equation}
    R_{\rm V}^{\rm A}  = \frac{4.68\times10^{23}}{M_{\rm tot}}\,
    \frac{\alpha^{'}}{\alpha}
    \left(\frac{\left[\text{keV} \right]}{m_{\rm V}} \right)
    \left(\frac{\sigma_{\rm pe}}{\left[\text{b} \right]} \right)\,
    \text{kg}^{-1}\,\text{d}^{-1}\,
\label{eq:R_vector_abs}
\end{equation}
where $M_{\rm tot}$ (g/mol) is the molar mass of the target material.
\noindent The ALPs and DPs dimensionless couplings to electrons are parametrized via $g_{\rm ae}$ and ${\alpha^{'}/}{\alpha}$, respectively. In particular, $\alpha'$ denotes the hidden sector fine structure constant and is related to the kinetic mixing strength $\kappa$ of DPs via $\alpha'=\alpha \kappa^2$ \cite{Bloch:2016sjj}. For absorption of DPs, the expression in Eq.~\eqref{eq:R_vector_abs} is only valid for $m_{\rm V}\!\gtrsim \!100$ eV where in-medium effects are negligible~\cite{Bloch:2016sjj,Hochberg:2016sqx}. Compared to the former \textsc{Gerda} publication, the rate constants of proportionality were recalculated. A more precise numerical value of 1.47 instead of 1.2 and 4.68 instead of 4 was obtained for ALPs and DPs, respectively. These estimates align with the numbers published in \cite{Aprile:2020tmw}.

\indent In this study, a second process  
has been included. This is the dark Compton scattering $\text{DM} + e^\text{-}\! \rightarrow\! e^\text{-} + \gamma$ causing the release of a photon and an electron with fixed energies. For a non-relativistic incident DM particle having an energy equal to $\omega \approx m_{\rm DM}$, the recoil energy $T$ of the electron  and the energy $\omega$' of the emitted photon are~\cite{PhysRevLett.128.191801}
\begin{equation}
    T = \frac{\omega^2}{2(m_{\rm e}+\omega)}~~~ {\rm and} ~~~\omega' = \sqrt{T^2+2 m_{\rm e} T}~. 
\label{eq:photon_en}
\end{equation}
\noindent Adapting rate formulas from \cite{PhysRevLett.128.191801}, the dark Compton interaction rate becomes 
\begin{equation}
    R_{\rm a}^{\rm C} = 
    f_{\rm a}^{\rm C}\,N_{\rm e}\, 
    \frac{1.27\times10^{24}}{M_{\rm tot}}\,
    g_{\rm ae}^2
    \left(\frac{\left[\text{keV} \right]}{m_{\rm a}} \right)
    \text{kg}^{-1}\,\text{d}^{-1}
\end{equation}
and
\begin{equation} 
    R_{\rm V}^{\rm C} = 
    f_{\rm V}^{\rm C}\,N_{\rm e}\,
    \frac{7.79\times10^{22}}{M_{\rm tot}}\,
    \frac{\alpha^{'}}{\alpha}
    \left(\frac{\left[\text{keV} \right]}{m_{\rm V}} \right)
    \text{kg}^{-1}\,\text{d}^{-1}\,,
\end{equation}
where $N_{\rm e}$ is the number of electrons of the target atom. The mass-dependent factors for ALPs and DPs are, respectively,
\begin{equation}
    f_{\rm a}^{\rm C}\left(m_{\rm a} \right) = \frac{m_{\rm a}^2\left(m_{\rm a}+2 m_{\rm e}\right)^2}{ \left(m_{\rm a} + m_{\rm e} \right)^4}\,
\end{equation}
and
\begin{equation}
    f_{\rm V}^{\rm C}\left(m_{\rm V} \right) = \frac{\left(m_{\rm V} + 2 m_{\rm e} \right) \left(m_{\rm V}^2+2 m_{\rm e}m_{\rm V} + 2m_{\rm e}^2\right)}{\left(m_{\rm V}+m_{\rm e} \right)^3}\,.
\end{equation}
As shown in~\cite{PhysRevD.104.083030} higher total interaction rates are expected for DM particle masses above $\sim$100~keV when including the dark Compton scattering process. 
In a realistic experimental environment, different scenarios are possible depending on the efficiency with which the final state particles are detected. The focus here is on events in which both the final electron and photon are detected within a single Ge detector, leading to a signal at energy $T+\omega'=m_{DM}$. The spectral shape of the signal in this absorption plus dark Compton scattering search is the same as in a pure absorption search, with the difference that the total expected signal is given by the sum of both contributions. 
\subsection{Nucleon decay}
\label{subsec:Nucleon decay} 
\begin{figure}[t!]
\includegraphics[width = 0.45\textwidth]{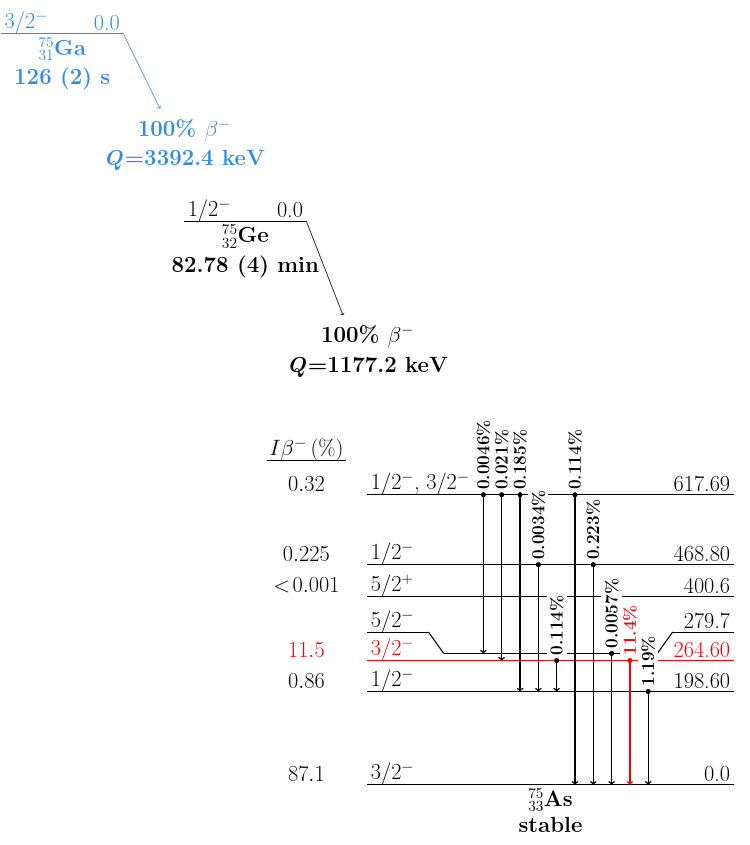}
\caption{Scheme of the $^{75}$Ge ground state $\beta$-decay to $^{75}$As and subsequent $\gamma$-decays, adapted 
from~\cite{NDS114:A=75}. The  $\beta$-decay (E$_\beta$\,=\,912.6 keV) to the second excited $^{75}$As state in coincidence with the 264.60~keV $\gamma$-ray is used to tag both the neutron and proton disappearance in $^{76}$Ge. 
Level and $\gamma$-ray of interest are highlighted in red.
The transition $^{75}$Ga$\rightarrow^{75}$Ge following $^{76}$Ge proton decays is shown in blue
}
\label{fig:75Ge_decay_scheme}
\end{figure}

Baryon and/or lepton number conservation violating single- and multi-nucleon decays are predicted in several extensions of the SM.
High nucleon decay lifetime sensitivities were already reached for light nuclei by tonne-scale experiments (see selected constraints listed in Sect.~\ref{sec:results:nucleon_decay}).
In this work, the inclusive, i.e. mode-independent, decay of a single neutron and proton in $^{76}$Ge is investigated. In the former, a neutron would disappear in a $^{76}$Ge nucleus, leading to an excited $^{75}$Ge nucleus if no particles other than photons are emitted. The energy release of approximately 9.4~MeV corresponds to the lowest nuclear separation energy for a nucleon in $^{76}$Ge~\cite{Huang:2021nwk,Wang:2021xhn}, which could then be observed. As in this energy release, neither the number of photons emitted nor their angular distribution is unique, the energy deposition in the \textsc{Gerda} detector array following such decay is difficult to model.
Hence, the subsequent low energy $\beta$-decay of the ground state $^{75}$Ge to an excited state of $^{75}$As, followed by a $\gamma$ de-excitation of the daughter nucleus, is considered. 
The dominant decay channel searched for in this analysis is the  $\beta$-decay to the $264.60$~keV level (E$_\beta$ = 912.6~keV, 11.5\% branching ratio), 
which is followed by the emission of a $264.60$~keV photon (see Fig. \ref{fig:75Ge_decay_scheme}). 
   
\indent The same method applies to the disappearance of a single proton. If a proton decays without the emission of accompanied nucleons, the produced $^{75}$Ga isotope undergoes $\beta$-decay to $^{75}$Ge with a half-life of 126(2)~s and a branching ratio of 100\%~\cite{NDS114:A=75}. Given that both neutron and proton decays can be probed with the coincident $^{75}$As 264.60~keV photon, this search is referred to as nucleon decay in the rest of the article. 
\\
\indent This study aims to establish limits for nucleon disappearance in $^{76}$Ge which has, to our knowledge, not yet been probed.

\subsection{Electron decay}
\label{subsec:Electron decay}

\begin{figure}[t]
    \centering
    \includegraphics[width = 0.49\textwidth]{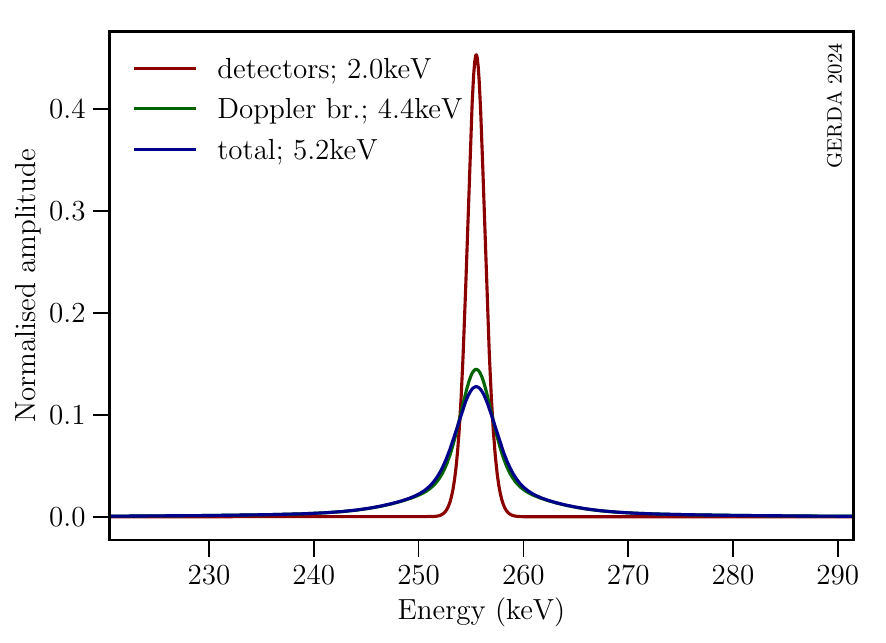}
    \caption{
    The contributions from detector resolution (red) and the Doppler-broadening (green) of lines from electron decay in the different atomic shells of germanium and argon (see Sect.~\ref{subsec:Signal model}). The total expected line shape of the electron decay signal at 255.9 keV is shown in blue. 
    All Gaussians are normalized to unit area.
    Indicated resolution values are given in FWHM
    }
    \label{fig:lineshape_256}%
\end{figure}

Many laboratory tests have been performed to test the fundamental U(1) gauge symmetry ensuring charge conservation (see selected constraints listed together with our results in Sect.~\ref{sec:results}, Table~\ref{tab:electron_exp_results}). The decay of an electron violating charge conservation could happen through the emission of three neutrinos, $e^\text{-}\!\rightarrow\! 3\nu_{\rm e}$, or a neutrino and a $\gamma$-ray, $e^\text{-}\!\rightarrow\!\nu_{\rm e}\gamma$. The former process has a maximal energy deposition that is equal to the maximal electron binding energy of $^{76}$Ge of $\sim$11.1~keV \cite{Lide:20042005}. As this value is below the trigger threshold of \textsc{Gerda}, this signature could not be used in this study. Instead, the decay $e^\text{-}\!\rightarrow \!\nu_{\rm e} \gamma$ was analysed. The peak is expected to lie around half of the electron mass, i.e. at $E_\gamma\sim255.5$~keV. In addition, the release of the relevant atomic binding energies causes both a Doppler broadening and a shift of the 255.5~keV peak for different electron atomic levels. 
In our setup electron decays could occur both within a germanium detector as well as in its surrounding materials which include neighboured germanium detectors and LAr.
If an electron decays within a detector's sensitive volume, both the photon energy and the one coming from the rearrangement of atomic shells, i.e. from X-rays or Auger electrons, are detected. Hence, for the \textit{i}-th atomic shell with binding energy $E_{\rm b,i}$, the total energy is
\begin{equation}
\label{eq:el:inner_decay}
    E_{\rm t,i}=\frac{m_{\rm e}-E_{\rm b,i}}{2}+E_{\rm b,i}=\frac{m_{\rm e}+E_{\rm b,i}}{2}~.
\end{equation}
In the case of an electron decaying outside the recording detector, 
the total detected energy equals  
\begin{equation}
\label{eq:el:outer_decay}
    E_{\rm t,i}=\frac{m_{\rm e}-E_{\rm b,i}}{2}~.
\end{equation}
Using Eq.~\eqref{eq:el:inner_decay} and the information provided in 
Sect.~\ref{sec:appendix_doppler_profile} of the Appendix the total energy recorded in a given germanium detector is expected to lie at 256.0 keV for electrons decaying within the detector's sensitive volume. Additionally, \textsc{Gerda} germanium detectors can detect outgoing photons coming from neighbouring germanium material undergoing the electron decay as well as from the surrounding LAr. Hence, using Eq.~\eqref{eq:el:outer_decay}, outgoing photons with energies of 255.0 keV and 255.3 keV, respectively, can be tagged. For each of these three contributions, the signal energy was derived as a weighted mean of energies $E_{\rm t,i}$ with the electron occupancy numbers as weights. 
Germanium and argon binding energies used in Eqs.~\eqref{eq:el:inner_decay} and~\eqref{eq:el:outer_decay} are listed in the Appendix (see Table~\ref{tab:lineshape_256_Ge_Ar} in Sect.~\ref{sec:appendix_doppler_profile}). The total signal energy is expected to be 255.9 keV by weighting for different source masses, electron occupancy numbers and detection efficiencies (see Eq.~\eqref{eq:doppler_line_shape_GeAr} in Sect.~\ref{sec:appendix_doppler_profile}).
Other surrounding materials. e.g. detector holders or electronic components, were not taken into account. 
Because of their low mass, they do not alter the results by more than a few percent.
The corresponding Doppler broadened line shape was determined as described in \cite{Klapdor-Kleingrothaus:2007aoa}. A discussion of the signal shape used in the present analysis is provided in the Appendix (see Sect.~\ref{sec:appendix_doppler_profile}). 
Figure~\ref{fig:lineshape_256} shows the 
final line shape, obtained by convolving the Doppler profile with a weighted Gaussian mixture distribution modelling the expected resolution broadening caused by the finite detector resolution (see Sect.~\ref{subsec:Signal model}). For the mixture model, the weights are defined as the exposures of each data set, separated by detector type and data-taking phase (see Sect.~\ref{sec:data_efficiency}). Considering the contributions of source detectors, surrounding detectors, and the LAr, the convolution yields a full width at half maximum (FWHM) of 5.2 keV, where the mixture model contributes 2.0 keV, and the full Doppler-broadened line 4.4 keV.

\section{Details of the \textsc{GERDA} experiment}
\label{sec:data_efficiency}

The \textsc{Gerda} experiment was located underground at the Laboratori Nazionali del Gran Sasso (LNGS) of INFN, in Italy, under the Gran Sasso mountain. The rock overburden offers a shield of about 3500~m water equivalent, reducing the cosmic muon flux by six orders of magnitude~\cite{Ackermann_2013}. 
Started in December 2015, the second phase of the experiment used 10 coaxial (Coax) detectors, 3 of them having a natural $^{76}$Ge isotopic abundance, together with 30 enriched Broad Energy Germanium (BEGe) detectors~\cite{GERDA:2017ihb}. 
In October 2017, the energy trigger threshold of detectors was lowered from $\mathcal{O}(100)$ to $\mathcal{O}(10)$ keV. 
Data taking was interrupted in April 2018 for a hardware upgrade by replacing one enriched Coax detector ($\sim$1 kg) and all natural Coax detectors by 5 new enriched inverted coaxial (IC) detectors, with a total mass of 9.6~kg~\cite{GERDA:2020xhi}.
Data taking was resumed in July 2018 and lasted until November 2019. Here, data collected before (after) the 2018 upgrade are referred as Phase~II (Phase~II+) data. 
HPGe detectors were arranged in 7 strings, each of them enclosed in a transparent nylon cylinder that mitigates the $^{42}$K background~\cite{Lubashevskiy:2017lmf}.
The 7-string array was operated inside a 64~m$^3$ LAr cryostat~\cite{Knopfle:2022fso} which provided both cooling and a high purity, active shield against background radiation. To detect scintillation light, the LAr volume around the array was instrumented with a curtain of wavelength-shifting fibers coupled to silicon photo-multipliers. Additionally, 16 cryogenic photomultiplier tubes (PMTs) were mounted on the copper plates at the two ends of the cylindrical LAr volume~\cite{GERDA:2017ihb,Janicsk_Cs_thy_2011}.
During the 2018 upgrade, the geometrical fiber coverage was improved with the addition of an inner curtain~\cite{GERDA:2020xhi}. 
The LAr cryostat was placed inside a tank containing 590~m$^3$ of ultra-pure water. The water tank was instrumented with 66 PMTs that help to detect Cherenkov light coming from muons passing through the experimental volume. The muon-induced background was further reduced to negligible levels by operating plastic scintillator panels placed on the roof of the clean room~\cite{Freund:2016fhz}.

\subsection{Data selection}
\label{subsec:Data selection}

In this paper, only Phase~II and II+ data collected after the installation of the LAr veto system \cite{GERDA:2017ihb} were considered. 
Different data sets were used for bosonic DM and particle disappearance searches. 
Table~\ref{tab:individual_exposure} shows the exposure levels evaluated for enriched Coax, BEGe and IC detectors, 
during different periods of data taking. Natural coaxial detectors were left out of the analysis because of their unstable 
behaviour that translated into low duty factors. Pulse shape discrimination (PSD) cuts, which had been optimised 
for the $0\nu\beta\beta$ decay search, were not applied in this study. 
Total exposure for all searches is 105.5~kg\,yr except for the bosonic DM search below 196 keV 
where it is 60~kg\,yr (see below).

\begin{table}[h!]
\caption{Exposures accumulated with indicated detector types during \textsc{Gerda} Phase~II (up to April 2018) and Phase~II+ (from July 2018). $R$ denotes the energy range of the respective spectra used for analysis in the bosonic DM search.
At the chosen energy bin size of 1 keV (see Sect.~\ref{subsec:Signal model})  
exposures for the energy intervals of $65\,-\,195$ keV and $196\,-\,1021$ keV are 
$\mathcal{E}_1$=60.0 kg\,yr and $\mathcal{E}_2$=105.5 kg\,yr, respectively. 
}
\begin{center}
\label{tab:individual_exposure} 
\begin{tabular}{lcccc}
\hline\noalign{\smallskip}
\textbf{Data collection} & \textbf{$\boldsymbol{R}$ (keV)} & \multicolumn{3}{c}{\textbf{Exposure (kg yr)}}  \\
\cmidrule(lr){3-5}  && \textbf{Coax} & \textbf{BEGe} & \textbf{IC} \\
\noalign{\smallskip}\hline\noalign{\smallskip}

    \text{Dec 2015\,-\,Oct 2017} & 196-1021 & 21.1 & 24.4 & -  \\
    \text{Oct 2017\,-\,Apr 2018} & 65-1021 & 7.5 & 8.4 & - \\
    \text{Jul 2018\,-\,Nov 2019} & 65-1021 & 13.2 & 22.2 & 8.7 \\
\noalign{\smallskip}\hline
\end{tabular}
\end{center}
\end{table} 
\noindent All searches share the same set of cuts, except the search for nucleon decay where the simultaneous firing of two detectors is required. This cut is henceforth referred to as the multiplicity 2 (M2) cut.
Quality cuts were applied to remove non-physical events starting from the inspection of waveform parameters.
Additionally, muon-induced events and events leading to energy depositions in the LAr were vetoed. 

\begin{figure}[t]
    \centering
\includegraphics[width = 0.49\textwidth]{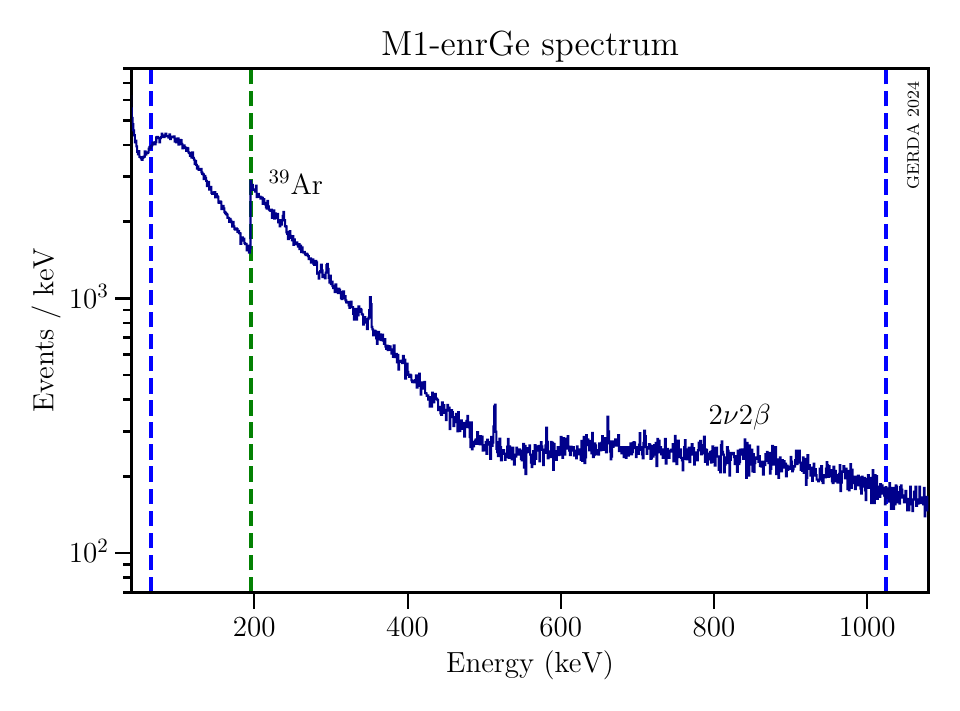} 
    \caption{Combined \textsc{Gerda} Phase~II/II+ spectrum of event multiplicity 1 after quality, 
muon veto, and LAr cuts.  
    The dominant background contributions from $^{39}$Ar $\beta$ decay and $^{76}$Ge 
$2\nu\beta\beta$ decay are indicated. The green dashed line separates the regions $65-195$ keV and 
$196-1021$ keV with exposure $\mathcal{E}_1=60.0$ kg\,yr and $\mathcal{E}_2=105.5$ kg\,yr, 
respectively (see Table~\ref{tab:individual_exposure}). The blue dashed lines mark the energy range 
inspected for bosonic DM candidates, i.e. 65-1021 keV}
    \label{fig:M1data}
\end{figure}

\paragraph{Bosonic dark matter}
A generic peak search was performed to look for signatures of a monoenergetic peak caused by the interaction of bosonic DM. The energy spectrum was filled only with events of multiplicity one (M1), i.e. events triggering only one Ge detector.
A histogram of the final M1 data set is shown in Fig.~\ref{fig:M1data}.
The bosonic DM analysis is performed in the interval 65(196)\,-\,1021~keV. The upper interval edge was fixed below $2m_{\rm e}$, the energy threshold of decays into electron-positron pairs. 
The lower energy bound was motivated by the analysis threshold of the Ge detector. Until October 2017, events were accepted if their energy exceeded $\geq\!195$ keV. Afterwards, the detector thresholds were lowered, thus, in addition, the data starting from 65 keV became available for this analysis. 
This change of thresholds causes the jump around 195~keV in the M1 energy spectrum of Fig.~\ref{fig:M1data}. 
More details are given in the Appendix (see Sect.~\ref{sec:appendix_bkgr}).
The $^{39}$Ar $\beta^-$ decay is well visible, up to the end-point energy of $565(5)$~keV~\cite{database1999}.
This $^{39}$Ar background is the reason why only full energy depositions were considered also for the dark Compton scattering process.  
Beyond $\sim$500~keV, the background continuum is dominated by the $^{76}$Ge two-neutrino double-beta ($2\nu\beta\beta$) decay characterized by an end-point energy of $Q_{\beta\beta}=2039.06$ keV~\cite{GERDA:2020xhi}. After applying the LAr cut, an almost clean $2\nu\beta\beta$ decay spectrum is observed (see Sect.~\ref{sec:bkgr}).

\begin{figure}[t]
    \centering
    \includegraphics[width = 0.49\textwidth]{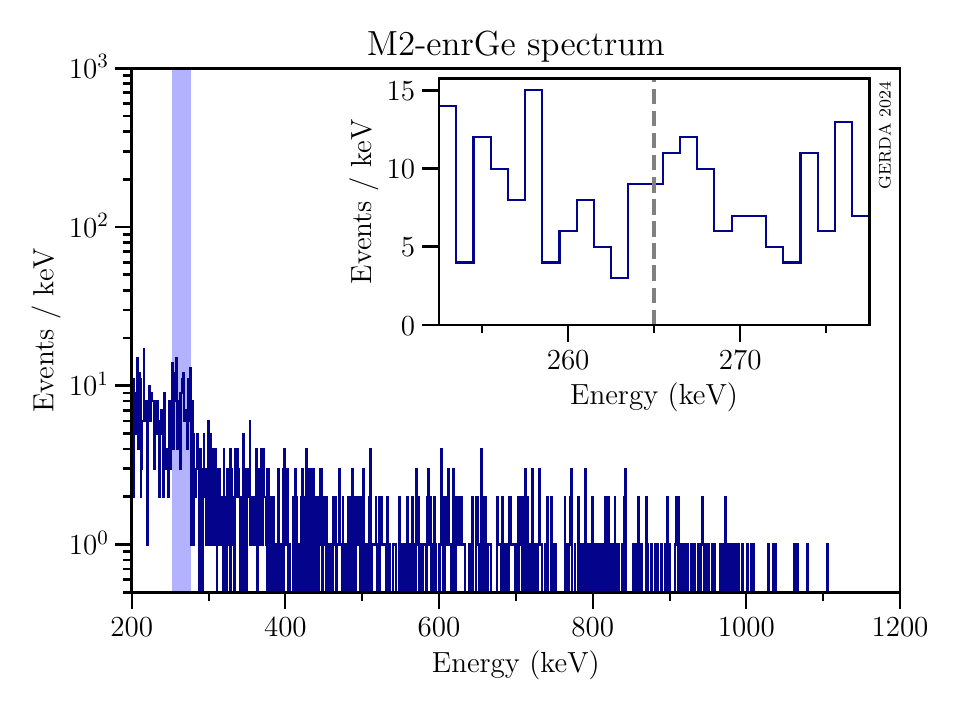}
\caption{
    Histogram of multiplicity 2 (M2) events; 
    see text for more details.
    The spectrum accounts for M2 events that survived quality cuts as well as muon and LAr vetoes. The inset shows the data in the $\gamma$-window (blue band) inspected for the nucleon decay signal, i.e. $E_\gamma\pm 12.5$ keV with $E_\gamma\sim265.0$~keV (gray dashed line) }
    \label{fig:M2data}
\end{figure}

\paragraph{Nucleon decay}
The study of a single nucleon decay in $^{76}$Ge was performed by searching for a $\beta$ particle with maximum energy \mbox{$E_\beta=912.6$~keV} and a coincident $\gamma$-ray of energy $E_\gamma=264.60$~keV (see Fig.~\ref{fig:75Ge_decay_scheme}).
The emitted $\beta$ particle is expected to be seen in the same detector where the nucleon decay happened since the range of an electron in germanium material is of $\mathcal{O}(10\mu\text{m}-1\text{mm})$ for the energy range
from 50 keV to 1 MeV~\cite{XCOM}. 
The photon may escape and propagate through the LAr to a neighbouring detector. Although the probability of this scenario is rather low, using this coincident tagging in two HPGe detectors strongly reduces the background.
In a M2 event with energies $\left(E_1,\,E_2\right)$ and $E_{1} + E_{2}\!<\!Q_\beta+2\cdot\text{FWHM}(Q_\beta)$, 
the partner with energy $E_{1(2)}$ is classified as $\gamma$ candidate if:
i)  $E_{2(1)}\!<\!E_\beta+2\cdot\text{FWHM}(E_\beta) \sim 918$ keV, 
or ii) $E_{1}$, $E_{2}$ are both within the $\gamma$-window and $\left|E_{1(2)}-E_\gamma \right| \!<\! \left|E_{2(1)}-E_\gamma \right|$. 
If both energies are outside the $\gamma$-window, arbitrarily the energy $E_{1,2}$ with the lower DAQ channel number is used to populate the M2 histogram.
Fig.~\ref{fig:M2data} shows the resulting  M2 histogram with the blue band indicating the $\gamma$-window, i.e. the region in which the search for the $^{75}$As de-excitation photon at 264.60 keV is performed: a $\pm$ 12.5 keV wide window around $E_\gamma = 265$~keV.
The width of this fit window was chosen sufficiently large both to contain the potential signal and to correctly model the background with a 1st order polynomial. Note that the choice made when $E_{1}$ and $E_{2}$ are both outside the  $\gamma$-window has no effect on the nucleon-decay analysis that focuses on events within the  $\gamma$-window. More details on the signal model and the systematic uncertainties related to the choice of the search window width are given in Sect.~\ref{subsec:Signal model} and~\ref{subsec:systematics}, respectively.
\paragraph{Electron decay}
For the analysis of the electron decay into $\nu_{\rm e}\gamma$, a broadened $\gamma$-line signal has to be considered (see Sect.~\ref{subsec:Electron decay}). Limiting the analysis to full energy $\gamma$ peaks, the same M1 data set was used as for the bosonic DM analysis.

\subsection{Detection efficiencies}
\label{subsec:Detection efficiencies}
To estimate the expected detection efficiencies, simulations were run in the MAjorana-GErda (\textsc{MaGe}) framework~\cite{Boswell:2010mr}. \textsc{MaGe} is a GEANT4-based software tool that allows users to generate simulated background and signal histograms for the \textsc{Gerda} experiment. Separately for each detector type (Coax, BEGe, and IC), three different sets of particle emissions ($e^\text{-}$, $\gamma$, $e^\text{-}+\gamma$) were simulated, as well as $^{75}$Ge decays. 
For all simulations, a set of $10^7$ primary particles was generated, uniformly distributed over the detector array.
Details on the simulation settings are reported in the following paragraphs. The generated raw files provide several pieces of information, e.g., the positions of the primary vertex, the hit energy depositions, and the particle types. The simulated events were then processed, 
taking into account specific settings for each experimental run, e.g., trigger thresholds, switched-off detectors, 
and dead layer models~\cite{Lehnert:2016}.
Acceptance efficiencies for the muon veto together with the quality cuts and the LAr veto were obtained as exposure-weighted averages of Phase~II and~II+ efficiencies~\cite{GERDA:2020xhi}. For a given cut, the total acceptance efficiency is 
\begin{equation}
    \epsilon_{\rm cut} = \frac{1}{\mathcal{E}}\left( \epsilon_{\rm cut,\,II} \cdot \mathcal{E}_{\rm II} +  \epsilon_{\rm cut,\,II+} \cdot \mathcal{E}_{\rm II+} \right)\,.
\end{equation}
Using exposures $\mathcal{E}_{\rm II}=61.4$\,kg yr and $\mathcal{E}_{\rm II+}=44.1$\,kg yr, total cut efficiencies of $\epsilon_{\mu}=0.999(1)$ and $\epsilon_{\rm LAr}=0.979(1)$ were obtained for the muon and LAr veto, respectively.
The total detection efficiency for a given final state $x$ is computed as 
\begin{equation}
\label{eq:tot_eff_weight}
    \epsilon_{x} = \epsilon_{\mu}\cdot\epsilon_{\rm LAr}\cdot \sum_{\rm i=1}^{N_{\rm d}} \frac{\mathcal{E}_{\rm i}\cdot\epsilon_{x\rm ,i}}{\mathcal{E}}\,,
\end{equation}
where $\mathcal{E}_{\rm i}$ and $\epsilon_{x,i}$ are the exposure and the 
efficiency for detector i and data set x, respectively. $N_{\rm d}$ denotes the total number of data sets. 
The full exposure $\mathcal{E}$ was divided into five data sets: enr-BEGe (32.8 kg\,yr) and enr-Coax (28.6 kg\,yr) from Phase~II, 
plus enr-BEGe (22.2 kg\,yr), enr-Coax (13.2 kg\,yr) and enr-IC (8.7 kg\,yr) from Phase~II+.
Table~\ref{tab:effi} provides a summary of the total detection efficiencies $\epsilon_{\rm X}$ for the potential signals in our search  
for new physics. 
More details are given below for each simulated process. 
For all simulated efficiencies, the statistical uncertainty is negligible given the high number of simulated events. 
The dominant systematic uncertainties affecting the efficiencies are the detectors' active volume uncertainties. 
For the nucleon decay search, there is an additional systematic uncertainty coming from the $^{76}$Ge enrichment level uncertainty. 
Systematic uncertainties are further commented in Sect.~\ref{subsec:systematics}. Summing in quadrature all contributions, 
a total uncertainty of 5\% is accounted in all searches.

\begin{table}[h!]
\caption{Summary of total detection efficiencies for indicated searches of potential signals from new physics. Quoted uncertainties include a total systematic uncertainty of 5\%; the statistical contributions can be neglected given the high number of simulated primaries}  
\begin{center}
\label{tab:effi} 
\begin{tabular}{lr}
\hline\noalign{\smallskip}
{\bf Bosonic DM} &\\
\hline
electron, $\epsilon_{e^\text{-}}$ & \\ 
~~ 65 keV & 0.852\,$\pm$\,0.043 \\
~~ 1021 keV & 0.805\,$\pm$\,0.040 \\
electron \& photon, $\epsilon_{e^\text{-}\land\gamma}$ & \\ 
~~ 65 keV & 0.839\,$\pm$\,0.042 \\
~~ 1021 keV & 0.165\,$\pm$\,0.008 \\
\noalign{\smallskip}
\hline\noalign{\smallskip}
{\bf Nucleon decay via $^{75}$Ge decay} &\\
\hline
coincidence of electron &                       \\
\& 264.60 keV photon, $\epsilon_{\rm n}$ & 0.0020\,$\pm$\,0.0001 \\
\noalign{\smallskip}
\hline\noalign{\smallskip}
{\bf Electron decay} &\\
\hline
$m_{\rm e}$/2 keV $\gamma$-ray emitted & \\ 
~~within recording detector, $\epsilon_{\rm Ge, det}$ & 0.419\,$\pm$\,0.021 \\
~~by neighbouring Ge material, $\epsilon_{\rm Ge, mat}$ &  0.034\,$\pm$\,0.002 \\
~~by LAr, $\epsilon_{\rm Ar}$ &  0.00070\,$\pm$\,0.00004 \\
\noalign{\smallskip}\hline
\end{tabular}
\end{center}
\end{table} 

\paragraph{Bosonic DM}
Simulations of electron energies in the interval 
65 to 1021 keV are required for the bosonic DM absorption channel, while for the dark Compton scattering channel the simulation of 
electrons and photons in the final state
is needed. Starting at 65 keV, efficiencies were computed as the ratio between the number of events in the full-energy peak and the number of 
simulated particles in steps of 1 keV. 
Primaries were simulated separately for each phase (Phase~II or Phase II+) and detector type. The total detection efficiencies were calculated as 
exposure-weighted means for the entire data-taking time and overall detector types (see Eq.~\eqref{eq:tot_eff_weight}). 
Including acceptance efficiencies for quality cuts, muon veto and LAr veto, total detection efficiencies for tagging electrons range from 
0.852$\,\pm\,$0.043 at 65~keV to 
0.805$\,\pm\,$0.040 at 1021 keV.
The same energy grid was used for the total energy when generating electrons plus photons from a single vertex with the energy constraints  
given by Eq.~\eqref{eq:photon_en}. Including all cuts, total detection efficiencies for tagging simultaneously electrons and photons at energy $T+\omega'=m_{DM}$ range from 0.839$\,\pm\,$0.042 
at 65~keV to 0.165$\,\pm\,$0.008 
at 1021 keV. At higher energies, the efficiency rapidly decreases because the probability of losing photons gets higher. In the window 65-1021 keV, the $\gamma$ attenuation length in Ge material ranges from $\mathcal{O}(\text{mm})$ up to $\mathcal{O}(\text{few cm})$ for energies above $\sim100$ keV~\cite{PhysRevLett.128.191801,osti_76335}. 
Escaping photons deposit energy either outside Ge material (if in LAr, the full event is discarded), leading to electron only signals at energy $T\!<\!m_{DM}$, or in a second germanium detector, leading to M2 events that are discarded from the bosonic DM analysis.
\paragraph{Nucleon decays via $^{75}$Ge}
Applying the same energy cuts used for building the M2 data set (see Sect.~\ref{subsec:Data selection}),
the $\beta$ decay of $^{75}$Ge and the subsequent gamma decays in $^{75}$As were simulated as well. 
Weighting over individual data sets with their exposures, a total detection efficiency of 0.0020$\,\pm\,$0.0001 
was derived. 
\paragraph{Electron decay}
The detection efficiency of measuring a $\sim256$~keV photon released after the electron decay in the Ge detectors and LAr volume was separately simulated. 
The efficiency, averaged over the exposure and accounting for the applied cuts, is found to be 0.419$\,\pm\,$0.021 for decays recorded in germanium detectors and 0.034$\,\pm\,$0.002 
for decays originating from detectors surrounding the one that fully recorded the outgoing photon. 
The efficiency of tagging photons originating in LAr was found to be 
(7.0\,$\pm$\,0.4)$\times$10$^{-4}$. 
This contribution was simulated in a cylinder with a radius of 0.8 m and a height of 1.4 m shielding the detector array, for a total mass of $m_{\rm Ar} = 3884.1$~kg. 

\section{Analysis methods}
\label{sec:methods}

\subsection{Signal model}
\label{subsec:Signal model}
In all signal channels searched for, full energy depositions within the Ge detectors are assumed, leading to peaks 
above the background continuum. The expected line at a probed energy would be constrained by the finite energy resolution of the detectors. The signal shape was thus modelled as a Gaussian profile under the assumption of a symmetric line shape for full charge collections. In the case of the electron decay channel, the line would be further broadened because of the Doppler effect as described in Sect.~\ref{subsec:Electron decay}. 
Given that all data were merged over different detector channels, the signal shape was a mixture of individual Gaussian distributions for each detector. 
The energy resolution (in standard deviations of a Gaussian peak) within different detector types operated in \textsc{Gerda} agree very well on the order of $\mathcal{O}(1~\text{keV})$, with systematic uncertainties of approximately 0.1-0.2 keV, which comply with the systematic uncertainty on the energy scale~\cite{GERDA:2021pcs}. The exposure-weighted resolution $\sigma$ ranges from 0.9~keV up to 1.2~keV in the bosonic DM interval of interest of 65~keV to 2$m_{\rm e}$. For particle disappearances at $\sim$265~keV and $\sim$256~keV, the energy resolution $\sigma$ is 0.9~keV. 
A bin size of 1~keV was thus chosen, being the closest integer to the energy resolution in standard deviations. Compared to this width, the uncertainties mentioned above are sufficiently small to accurately model the peak shape via a Gaussian mixture model over detector types, instead of using a full mixture model over all individual detector channels. The weights in the mixture model are the exposures of the individual detector types, as well as the two data-taking phases. Both signal centroid and resolution, as measured from approximately weekly calibrations~\cite{GERDA:2021pcs}, were fixed for every probed signal model, leaving only the signal strength amplitude as a free parameter in the signal shape to be fitted.\\
\indent For a DM signal model, the search window was limited to 25~keV, centred at the incoming DM mass particle, which is sufficiently large to compare the potential signal with $\sim$1~keV resolution in standard deviations to the wide background continuum discussed below. Every integer mass value in the search range of 65-$2m_{\rm e}$~keV was probed iteratively. For the nucleon decay, the same search window width was used but evaluated for the coincident M2 data centred at $E_\gamma\sim265$ keV. For the electron decay channel, owing to the broadening, the search window was increased to a width of 120~keV, ranging from 196~keV to 316~keV.

\subsection{Background model}
\label{sec:bkgr}
\paragraph{Background continuum}
The \textsc{Gerda} background model after the LAr veto cut does not fully cover the energy range of interest~\cite{GERDA:2022hxs}. Hence, it does not reproduce the observed $^{39}$Ar dominated spectral shape at lower energies. 
Thus, an empirical fit model, motivated by the underlying physical processes, was applied to constrain the background continuum in the M1 data set. 
The $2\nu\beta\beta$-decay dominated upper half of the signal range was modelled with a polynomial function. The dominating $^{39}$Ar $\beta$-decay background contribution at energies below approximately 500~keV was modelled with a modified $\beta$-decay distribution~\cite{Awodutire:2021,Nadarajah2014}. 
Owing to the propagation of the emitted electrons through the cryogenic liquid, resulting in strong bremsstrahlung emissions, a modification to the original $\beta$-decay shape was needed. 
Plots of the empirical background model as applied for the signal search, and an evaluation of its accuracy to describe the data, are provided in the Appendix (see Sect.~\ref{sec:appendix_bkgr}). \\
\indent No background decomposition of the M2 energy spectrum shown in Fig.~\ref{fig:M2data} is available. 
These events have a different energy distribution compared to M2 data shown in \cite{GERDA:2019cav}. The difference comes from having applied both an energy cut to M2 events and the LAr veto in this paper. 
Moreover, the M2 spectrum used in \cite{GERDA:2019cav} contains the sum of the two coincident energies. 
The $\gamma$ energy spectrum was instead fitted with a linear function of energy in a 25 keV wide interval around the expected signal 
at $\sim$265.0 keV. 

\paragraph{$\gamma$-ray background}

Background $\gamma$-radiation emitted
from surrounding materials creates the very same peak profile in the data as the bosonic DM signals searched for. 
Thus, the $\gamma$-lines cannot be distinguished from these signals.
Hence, as a first step, a generic search for any peak-like excess above the background continuum was performed, independently of whether an excess was caused by a known isotope transition or new physics. If the significance of an excess exceeded 3$\sigma$, and if it could be explained by a known $\gamma$-transition,
the corresponding $\gamma$-line peak was added to the background model. When evaluating limits on the bosonic DM interactions and the electron decay lifetime, the background model function was refitted in a second step, including the $\gamma$-rays identified during the generic search. When determining bosonic DM limits, the $\gamma$-line peak energies were excluded together with 3 bins on the right and on the left, corresponding to an exclusion window of approximately 2.5~FWHM width for each detected $\gamma$ line. 

\subsection{Statistical frameworks}
\label{subsec:Statistical frameworks}

Two independent statistical analyses were conducted to identify a potential excess at any probed energy value. A binned Bayesian fit of the signal peak above the background model was performed in the respective signal window, employing a positive uniform prior for the signal strength amplitude.
In addition, a Frequentist fitting procedure was employed using the profile likelihood-ratio test statistics from \cite{Cowan:2010js}. Asymptotic distributions were assumed to hold, and the physically allowed signal strength was constrained to the positive domain. Both statistical approaches are described in more detail in the Appendix (see Sect.~\ref{sec:appendix_statistical_frameworks}). In both methods, a 3$\sigma$ threshold was required to identify an indication of a potential signal. A 4$\sigma$ effect was required to claim signal evidence in the particle decay searches, a 5$\sigma$ effect in the bosonic dark matter search which is prone to a strong look-elsewhere effect as discussed in Sect.~\ref{sec:appendix_statistical_frameworks}.
\begin{figure}[t]
    \centering
    \includegraphics[width = 0.49\textwidth]{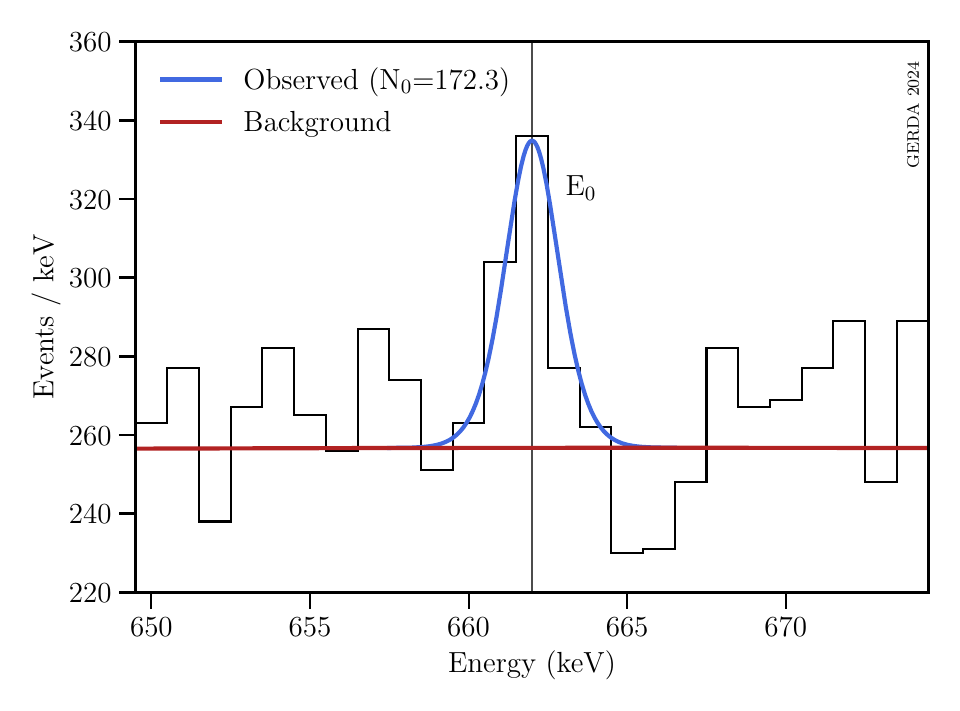} 
    \caption{Part of the M1 spectrum shown in Fig.~\ref{fig:M1data} with an example of a Bayesian fit at 662~keV (vertical line). The empirical background contribution is shown in red, while the best-fit model is shown in blue. $N_0$ denotes the best-fit signal strength. The signal excess of 5.1$\sigma$ can be explained by the 661.7~keV $\gamma$-line from $^{137}$Cs (see Table~\ref{tab:detected_lines})}
    \label{fig:fit_plot_DM}%
\end{figure}
An example of a Bayesian fit is shown in Fig.~\ref{fig:fit_plot_DM} at the potential mass of 662~keV for which an excess of 5.1$\sigma$ has been observed and attributed to the known $^{137}$Cs line at $\sim$662.0~keV. The observed local p-values for each probed peak position in the bosonic DM search range, as determined in the Frequentist framework, are shown in Fig.~\ref{fig:p-values}.
Overall, nine expected $\gamma$-ray transitions were identified, plus one unknown excess at 710~keV, as listed in Table~\ref{tab:detected_lines}. The global significance of the unidentified excess is discussed in the Appendix (see Sect.~\ref{sec:appendix_statistical_frameworks}). As the corresponding local significance of this peak remains below the evidence threshold, it was concluded that no bosonic DM signal was found.  

\begin{figure}
    \centering
    \includegraphics[width = 0.49\textwidth]{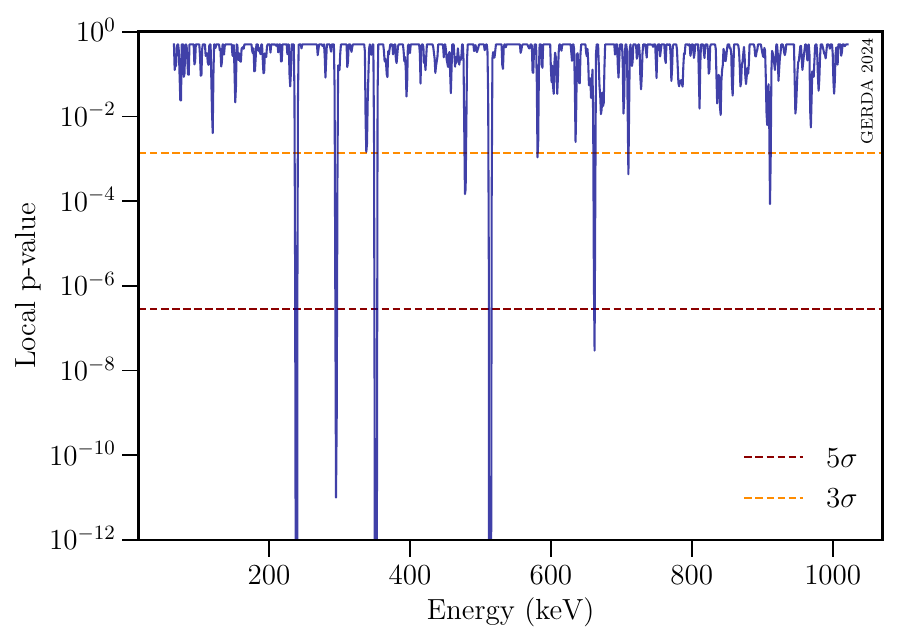} 
    \caption{Plot of the local p-values of all count strength amplitudes versus the tested energies for the DM search. 
    Apart from the  $3\sigma$ excess at 710~keV all other local excesses with $\geq\!3\sigma$ can be attributed to known $\gamma$ transitions (see Table~\ref{tab:detected_lines}) }
    \label{fig:p-values}%
\end{figure}

\begin{table}[h!]
\caption{List of energy ranges $R$ where $\geq\!3\sigma$ excesses are found by the Bayesian and/or Frequentist fits, and their maximum significance $S$ (Bayesian, Frequentist).  The most likely origin of these peaks are $\gamma$ transitions from indicated nuclei; the respective energies $E_\gamma$ are taken from \cite{reference.wolfram_2022_isotopedata}}
\begin{center}
\footnotesize{
\centering
\begin{tabular}{cccc} 
 \hline
 \textbf{$\boldsymbol{R}$ (keV)} & $\boldsymbol{S}$ $\boldsymbol{(\sigma)}$ & \textbf{Origin} & $\boldsymbol{E_\gamma}$ \textbf{(keV)} \\ [0.5ex] 
 \hline
237 - 240 & 8.4, 8.5 & $^{212}$Pb & 238.632\,(2)\\
293 - 297 & 6.4, 6.7 & $^{214}$Pb & 295.224\,(2)\\
338 & 2.9, 3.0 & $^{228}$Ac & 338.320\,(5)\\
349 - 353 & 10.0, 10.7 & $^{214}$Pb & 351.932\,(2)\\
477 - 479 & 3.6, 3.6 & $^{228}$Ac & 478.4\,(5)\\
512 - 516 & 8.8, 10.2 & $^{85}$Kr & 513.997\,(5)\\
581 & 3.1, 3.1 & $^{208}$Tl & 583.187\,(2)\\
660 - 663 & 5.1, 5.4 & $^{137}$Cs & 661.657\,(3)\\
710 & 2.9, 3.3 & - & -\\
910 - 912 & 3.5, 3.8 & $^{228}$Ac & 911.196\,(6)\\
 \hline
\end{tabular}
\label{tab:detected_lines}
\centering}
\end{center}
\end{table}

\noindent Also for the nucleon and electron decay channels no significant signal excess was seen. Hence, upper limits were evaluated for all new physics searches independently at 90\% CI and 90\% CL (see Sect.~\ref{sec:appendix_statistical_frameworks} for technical details). The corresponding sensitivities were determined via Monte Carlo (MC) simulations in the Bayesian case, and via Asimov data sets~\cite{Cowan:2010js} in the Frequentist method.

\subsection{Systematics}
\label{subsec:systematics}
Different sources of systematic uncertainties were investigated. In the Bayesian framework, the accuracy of expected limits was checked via MC simulations. At each probed energy value, $10^3$ toy-MC spectra were generated assuming no signal and Poisson fluctuations for the number of background events. 
Each toy spectrum was fitted with a signal+background model. The distribution of the derived limits for the signal strength amplitudes was used to derive the median sensitivity. 
Measured limits are well contained within the simulated expectation bands and agree with the median sensitivity expected in case of no signal (see Fig.~\ref{fig:dark_alps_gerda_comparison_stats_sensitivity} in Appendix~\ref{sec:appendix_bDM_comparison_stats}).
In the Frequentist case, the Asimov data sets were employed to investigate systematic uncertainties. Here both the accuracy of the Asimov sensitivity estimations and the assumption of asymptotic distributions for the limit evaluation were confirmed via $10^6$ MC simulations at the equally spaced energies $\lbrace 100, 150, ...,$$\,1000 \rbrace$~keV for bosonic DM searches and at the energies of the nucleon and electron decay channel. The resulting uncertainties are within 11 (3)\% for the M2 (M1) data set, which is judged sufficiently accurate.\\
\noindent The systematic uncertainty on the bosonic DM results caused by the background modelling approach was checked via a different background fit. The results obtained with the empirical background fit model were compared to those obtained with a polynomial background continuum fit in each individual search window, in exact analogy to our former work shown in \cite{GERDA:2020emj}. The respective sensitivities reveal a systematic uncertainty of $\sim$1\%, indicating a good accuracy of the background modelling procedure. Here, the uncertainty was estimated as the median of all deviations between the two approaches. Following the same fitting treatment as in our previous work would change the Bayesian (Frequentist) limits by approximately 1 (2)\%, again estimated as the median deviation. 
\\ \noindent The impact of modelling the background continuum on the results for the electron (nucleon) decay channel was probed as well, using a second (first)  order polynomial function and different search window widths. 
The differences in the Bayesian (Frequentist) sensitivities for different fitting strategies remain within approximately 2 (4)\% for the nucleon decay analysis and are $\sim$1\% for the electron decay search. \\

\noindent Furthermore, the effect of the bin width has been investigated. Probing bin widths within reasonable proximity to the energy resolution scale in standard deviations of 1~keV, with a systematic uncertainty of around 0.1-0.2~keV, reveals an uncertainty on both bosonic DM results of $\sim$7\%. The uncertainties are slightly smaller for the decay channel sensitivities, independently of the statistical framework.\\ \noindent The detector-geometry-related uncertainties caused by the active volume or the level of enrichment in $^{76}$Ge (the latter being relevant for the nucleon decay search only) have an impact of approximately 4\% and 2\%, respectively. These were estimated as the exposure-weighted mean of the active volume and enrichment fraction uncertainties of the different detector types~\cite{GERDA:2020xhi}.

\section{Results}
\label{sec:results}

\subsection{Bosonic dark matter}
\label{sec:results:bDM}
No evident excess caused by bosonic DM interactions has been found beyond the expected fluctuations of the continuous background. Using the interaction rate formulas shown in Sect.~\ref{subsec:Bosonic dark matter}, the derived count strength limits $N_{\rm up}$ at 90\%~CI and CL are converted into upper limits on the maximal physical interaction strength of ALPs and the kinetic mixing of DPs. In particular, the conversion formula reads
\begin{equation}
\begin{aligned}
    g_{\rm \phi} = \frac{N_{\rm up}}
    {\mathcal{E_{\rm 1(2)}}\cdot 365.25 \cdot R_{\rm \phi}}\,,
\end{aligned}
\label{eq:R_converted}
\end{equation}
where $\phi$ denotes the DM candidate of interest, which can either be an ALP ($\phi \equiv a$ and $g_{\phi}\equiv g_{\rm ae}^2$) or a DP ($\phi \equiv V$ and $g_{\phi}\equiv {\alpha^{'}/}{\alpha}$), and  $\mathcal{E_{\rm 1(2)}}$ 
the exposure of 60.0 or 105.5 kg\,yr (see Table~\ref{tab:individual_exposure}). The total DM interaction rate $R_{\rm \phi}$ ($\text{kg}^{-1}\text{d}^{-1}$) accounting for detection efficiencies shown in Table~\ref{tab:effi} is given by
\begin{equation}
    R_{\rm \phi} = \epsilon_{e^\text{-}} \cdot {R_{\rm \phi}^{\rm \,A}} + \epsilon_{e^\text{-}\land\gamma}\cdot R_{\rm \phi}^{\rm \,C}\,.
\label{eq:tot_rate}
\end{equation}
When computing the absorption interaction rates through Eqs.~\eqref{eq:R_pseudo_abs} and~\eqref{eq:R_vector_abs}, the photoelectric cross-section $\sigma_{\rm pe}$ for germanium target material was taken from Ref.~\cite{XCOM}. The molar mass $M_{\rm tot}=75.66$ g/mol of enriched Ge detectors was computed as 
\begin{equation}
    M_{\rm tot} = f_{\rm ^{76}Ge}\cdot M_{\rm ^{76}Ge} + \left(1-f_{\rm ^{76}Ge} \right)\cdot M_{\rm res}\,,
\end{equation}
where the \textsc{Gerda} exposure-weighted $^{76}$Ge enrichment fraction is $f_{\rm ^{76}Ge}=87.5\%$~\cite{GERDA:2020xhi}.
The molar mass of all isotopes but 76 present in enriched Ge detectors is computed as
\begin{equation}
    M_{\rm res} = \sum_{\rm i \neq 76} \frac{M_{\rm i}\cdot f_{\rm i}}{f_{\rm tot}}\,,
\end{equation}
for Ge isotopes $i=\{70,\,72,\,73,\,74\}$. Molar masses $M_{\rm i}$ are taken from \cite{XCOM}, while relative isotopic composition values $f_{\rm i}$ were taken from Table 1 of \cite{GERDA:2020xhi}, with $f_{\rm tot}=\sum_{\rm i \neq 76}f_{\rm i}$. 
In particular, $M_{\rm ^{76}Ge}=75.92$ g/mol and $M_{\rm res}=73.86 $ g/mol.
The derived limits on the kinetic mixing strength of DPs and the ALP-electron coupling are compared to other experimental results in Fig.~\ref{fig:dark_alps}. Constraints for specific masses are listed in the Appendix, see Table~\ref{tab:bDM_example_results} in Sect.~\ref{sec:appendix_bDM_comparison}. 
The results obtained with the Frequentist method largely align with the Bayesian results, but are slightly more stringent at the locations of underfluctuations below the expected background levels. In the Appendix, individual effects of the absorption and the scattering process on the total results are shown (see Sect.~\ref{sec:appendix_bDM_comparison}), and  the sensitivities compared as determined with the two different statistical approaches (see Sect.~\ref{sec:appendix_bDM_comparison_stats}).
\begin{figure*}[h]
\centering
    \includegraphics[width = 0.49\textwidth]{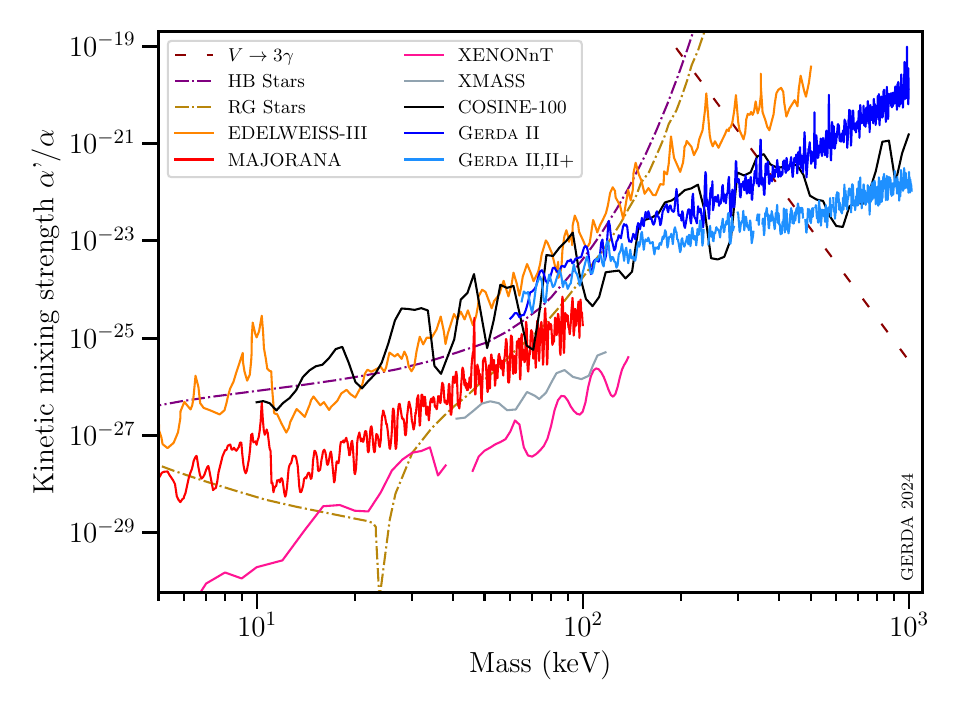} %
    \includegraphics[width = 0.49\textwidth]{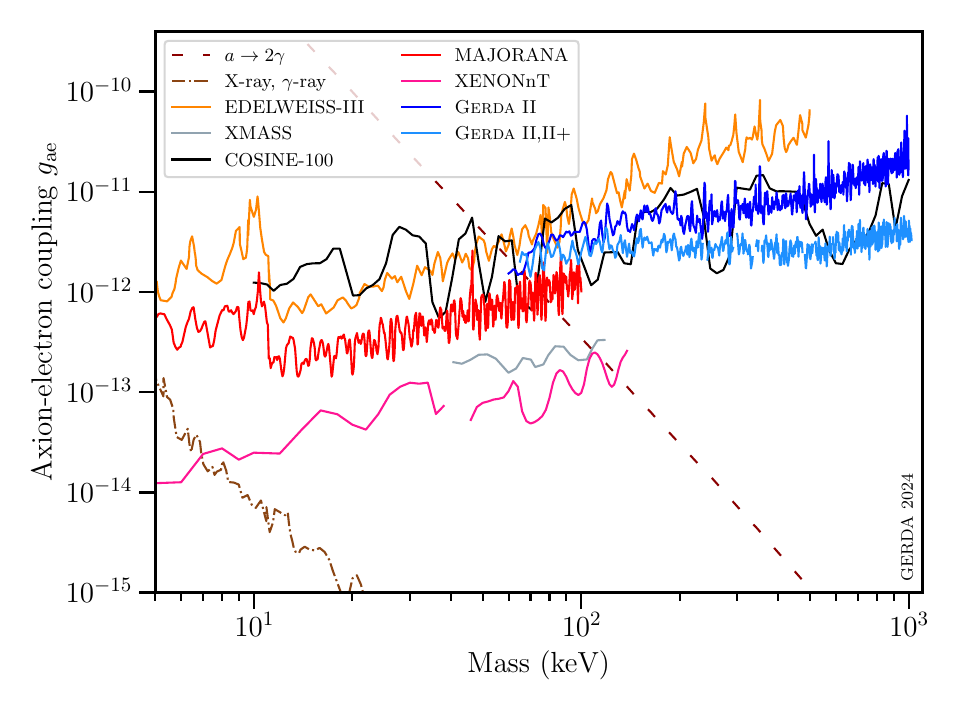} %
    \caption{Bayesian exclusion limits on bosonic DM couplings to electrons  obtained from \textsc{Gerda} Phase~II and Phase~II+ data (light blue line). The limits were deduced by converting the upper count strength limits into physics constraints including in the interaction rate both the photoelectric-like absorption and the dark Compton scattering processes, see Eq.~\eqref{eq:R_converted}. 
The regions around identified $\gamma$-lines (see Table~\ref{tab:detected_lines} and numerical data in Supplemental Material \cite{xyzdata}) have been omitted. 
    Left: Bayesian constraints at 90\% CI on the kinetic mixing strength of DPs. Right: Bayesian constraints at 90\% CI on the coupling strength of ALPs to electrons. Results from other direct detection experiments~\cite{PhysRevD.98.082004,PhysRevLett.132.041001,XENON:2022ltv,Sato:2020ebe,COSINE-100:2023dir} are shown, as well as the previous \textsc{Gerda} limits~\cite{GERDA:2020emj}. Note that in the COSINE-100 paper \cite{COSINE-100:2023dir} the previous numerical factors of 1.2 and 4 have been used in eqs. 4 and 5. The dashed, dark red line indicates the region below which the interpretation as a DM candidate being stable on the scale of the age of the Universe is valid without further assumptions~\cite{PhysRevLett.128.221302}. Indirect constraints from X-ray and $\gamma$-ray observations taken from Refs.~\cite{PhysRevLett.128.221302,AxionLimits} are indicated by the dot-dashed, brown line. Constraints derived from red giant (RG, dot-dashed, gold line) and horizontal branch (HB, dot-dashed, purple line) star energy losses are discussed in \cite{Li:2023vpv}
    }%
    \label{fig:dark_alps}%
\end{figure*}

\noindent The new limits derived by \textsc{Gerda} are among the most stringent direct measurement results between $\sim$140~keV and $2m_{\rm e}$, if not the best. 
Better constraints are reported only for masses in the intervals of about 245-280 keV and 570-670 keV by COSINE-100~\cite{COSINE-100:2023dir}. Comparing old \cite{GERDA:2020emj} and new \textsc{Gerda} limits improvements of almost up to two orders of magnitude are achieved at energies above $\sim$500~keV for the DP channel due to the domination of the Compton cross-section versus the absorption cross-section. For ALPs, this corresponds to an improvement of almost one order of magnitude. 
At intermediate energies, the doubled exposure in combination with the combined effect of absorption and scattering leads to about 2 to 10 times more severe constraints, depending on the precise energy and the particle candidate. At lower energies, the new results improve only marginally upon the limits derived in \cite{GERDA:2020emj}. The small improvement in this region is mostly triggered by an approximately four times higher exposure, meaning an expected improvement by a factor of 2 only, as the dark Compton process does not contribute relevantly in this range. Hence, the sensitivities of xenon-based direct DM detection experiments could not be reached, due to the higher background level in our low energy range and the lower exposure.
\subsection{Nucleon decays}
\label{sec:results:nucleon_decay}
A lower constraint on the nucleon lifetime based on the observed upper limit on the event number $N_{\text{up,n}}$ is calculated as
\begin{equation}
    \tau_{\text{low}} = \epsilon_{\rm n} \cdot N_{\rm eff} \cdot  \frac{N_{\rm A}}{N_\text{up,n}}  \cdot \mathcal{E} \cdot \frac{ f_{\rm ^{76}Ge}}{M_{\rm tot}} \,
\label{Eq:lifetime_limit}
\end{equation}
where $\epsilon_{\rm n}$ is the efficiency to tag a coincident electron-photon pair (see Table~\ref{tab:effi} in Sect.~\ref{sec:data_efficiency}), $N_{\rm eff}$ is the effective number of particles which can undergo the considered decay, and $N_{\rm A}$ is the Avogadro's constant. $M_{\rm tot}$ (kg/mol) and $f_{\rm ^{76}Ge}$ are given in Sect.~\ref{sec:results:bDM}, while the exposure $\mathcal{E}=105.5$ kg\,yr is taken from  Table~\ref{tab:individual_exposure}.
As described in Sect.~\ref{subsec:Nucleon decay}, only one specific branch of the inclusive nucleon decay is considered, i.e. the one in which the nucleon decays from one of the most external nuclear shells with the de-excitation of the daughter nucleus by $\gamma$-emission only, without subsequent emission of other particles. Hence, it is necessary to know the effective number of decaying neutrons (protons) inside the parent $^{76}$Ge nuclei, whose decay could produce the specific daughter nucleus $^{75}$Ge ($^{75}$Ga). Following Refs.~\cite{Bernabei:2000xp,Bernabei:2006tw,Hazama:1994zz,EvansJr:1977zuj}, the effective number $N_{\rm eff} = 16$ (14) for neutrons (protons) was obtained by using the single-particle shell model with a modified Woods-Saxon potential~\cite{Woods:1954zz,Schwierz:2007ve}, and the set of parameters adjusted for $^{76}$Ge. The calculations were done with the shell-model codes KSHELL~\cite{Shimizu:2013xba} and CoSMo~\cite{volya_2016}
comparing, where possible, our full range of the sub-shell nucleon binding energies with the values obtained in Refs.~\cite{Suhonen:2008zz,Hirsch:2012uz}.\\ 
\begin{figure}
    \centering
    \includegraphics[width = 0.49\textwidth]{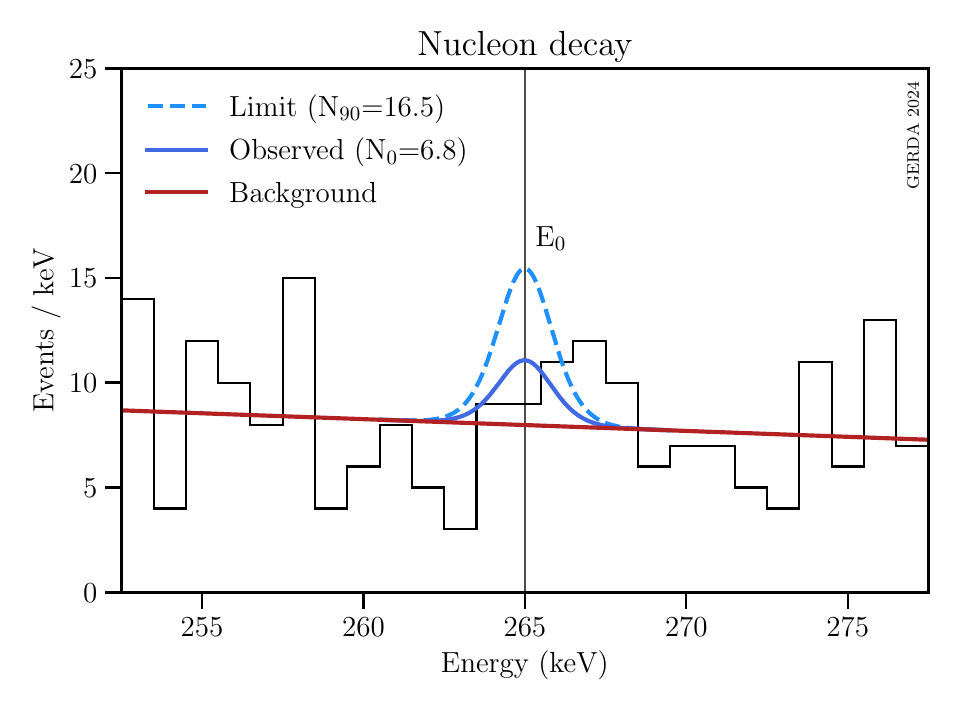} 
    \caption{Part of the M2 spectrum shown in Fig.~\ref{fig:M2data} with the Bayesian fit of the nucleon decay signal at $E_0\sim$265~keV. A 1st-order polynomial was used to model the continuous background
    }
    \label{fig:fit_plot_nucleon}%
\end{figure}

\noindent
In the Bayesian framework a best fit of 6.8 counts was obtained, with a significance of $1.1\sigma$ (see Fig.~\ref{fig:fit_plot_nucleon}). The $90$\%~CI upper limit is equal to $N_{\rm up,n}=16.5$ counts, 
and the median sensitivity is estimated to be $N_{\rm s,n}=10.5$ counts. In the Frequentist approach, the best-fit signal strength is 4.2 counts, corresponding to a significance of 0.7$\sigma$. This leads to a count limit of $N_{\rm up,n}=15.2$ counts with a median sensitivity estimate of $N_{\rm s,n}=9.8$ counts. 
\noindent The respective limits on the nucleon lifetimes estimated through Eq.~\eqref{Eq:lifetime_limit} are shown in Table~\ref{tab:lifetime_limits_recap}.
The lifetime limit for $N_{\rm eff} = 1$ is provided both as a measure of the inclusive {\it nuclear} decay rate and for comparison with other published limits, where different effective numbers of nucleons were used depending on the specific isotopes under consideration.

For a comparison with the results of previous nucleon disappearance studies see the detailed compilation of the Particle Data Group \lq p Mean Life\rq \ \cite{PDG2022.upd2023}. 
For inclusive decays of neutrons and protons bound in 
$^{129,136}$Xe~\cite{Bernabei:2000xp,Bernabei:2006tw} , $^{127}$I~\cite{Hazama:1994zz}  
and $^{130}$Te~\cite{EvansJr:1977zuj,Zdesenko:2003ph} mean life limits between 3.3$\times10^{23}$ and 8.6$\times10^{24}$ yr have been found. 
Orders of magnitude better limits are reported by the Borexino, KamLAND and SNO+ collaborations for the parent 
nuclei $^{12,13}$C~\cite{PhysRevLett.96.101802,Borexino:2003igu} and $^{16}$O~\cite{PhysRevD.105.112012} 
profiting from the huge mass of their low-background detectors. These latter experiments provide limits on the decay of bound nucleons 
into invisible modes where no energy is deposited in the detector in the decay itself. 
The best limits are provided by SNO+ for neutron and proton disappearance in $^{16}$O, 9$\times10^{29}$ yr and 9.6$\times10^{29}$ yr, 
respectively~\cite{PhysRevD.105.112012}.

\begin{table*}
\caption{Summary of results of the search for inclusive neutron ($n$) and proton ($p$) decays ($n,p\rightarrow{}X$) in $^{76}$Ge as well as for electron decay $e^\text{-} \rightarrow \nu_{\rm e} \gamma$.
For each decay, 
the observed best-fit value (${obs.}$) is shown together with its significance ($sig.$). 
The extracted upper limits at 90\% CI/CL and the median sensitivity for the signal strength are indicated with $N_{\rm up}$ and $N_{\rm s}$, respectively.
Lower lifetime limits (${L}$) on $\tau_{\text{low}}$ are deduced in the Bayesian and Frequentist frameworks according to Eqs.~\eqref{Eq:lifetime_limit}, \eqref{Eq:lifetime_limit_e} at 90\% CI and CL,
respectively, with the sensitivity $S$ equal to the median value assuming the background-only hypothesis. 
$N_{\rm eff}=16\,(14)$ denotes the effective numbers of neutrons\,(protons)  
used for deriving the nucleon lifetime limit. $N_{\rm eff}=1$ yields the corresponding nuclear decay rate limit.
As to electron decay, $N_{\rm eff}$ denotes the number of electrons in Ge and Ar atoms
}
\begin{center}
\label{tab:lifetime_limits_recap} 
\begin{tabular}{cccccccc}
\hline\noalign{\smallskip}
\textbf{Search} & \textbf{Framework} & \multicolumn{3}{c}{\textbf{Signal counts}} & $\boldsymbol{N_{\rm eff}}$ & \multicolumn{2}
{c}{$\boldsymbol{\tau_{\text{low}}}$\textbf{(yr)}}\\
\cmidrule(lr){3-5} \cmidrule(lr){7-8} \textbf{} & & \textbf{$\boldsymbol{obs.\,(sig.)}$}  & \textbf{$\boldsymbol{N_{\rm up}}$} & \textbf{$\boldsymbol{N_{\rm s}}$} & &
\textbf{$\boldsymbol{L}$} & \textbf{$\boldsymbol{S}$} \\
\noalign{\smallskip}\hline\noalign{\smallskip}
     $n,p\rightarrow{}X$ & Bayesian & 6.8 (1.1$\sigma$) & 16.5 & 10.5 & 1 & $9.1 \times 10^{22}$ & $1.4 \times 10^{23}$ \\
     & & & & & 16 ($n$) & $1.5 \times 10^{24}$ & $2.3 \times 10^{24}$ \\
     & & & & & 14 ($p$) & $1.3 \times 10^{24}$ & $2.0 \times 10^{24}$ \\
     & Frequentist & 4.2 (0.7$\sigma$) & 15.2 & 9.8 & 1 & $9.8 \times 10^{22}$ & $1.5 \times 10^{23}$ \\
     & & & & & 16 ($n$) & $1.6 \times 10^{24}$ & $2.4 \times 10^{24}$ \\
     & & & & & 14 ($p$) & $1.4 \times 10^{24}$ & $2.1 \times 10^{24}$ \\\hline
      $e^\text{-} \rightarrow \nu_{\rm e} \gamma$ & Bayesian & 15.3 (0.3$\sigma$) & 264.2 & 249.4 & 32 (Ge), 18 (Ar) & $5.4 \times 10^{25}$ & $5.7 \times 10^{25}$ \\
      & Frequentist & 3.8 (0.0$\sigma$) & 263.1 & 259.2 & 32 (Ge), 18 (Ar) & $5.4 \times 10^{25}$ & $5.5 \times 10^{25}$ \\
\noalign{\smallskip}\hline
\end{tabular}
\end{center}
\end{table*}
\subsection{Electron decay}
Similarly to Eq.~\eqref{Eq:lifetime_limit}, the constraint on the electron decay lifetime is calculated as
\begin{equation}
\begin{split}
    \tau_{\text{low}} = & ~\left(\epsilon_{\rm Ge, det} + \epsilon_{\rm Ge, mat}\right)  \cdot N_{\rm e, Ge}  \cdot  \frac{N_{\rm A}}{N_\text{up,e}}  \cdot   \frac{\mathcal{E}}{M_{\rm tot}} \\ & 
    + 
    \epsilon_{\rm Ar} \cdot N_{\rm e, Ar} \cdot \frac{N_{\rm A}}{N_\text{up,e}} \cdot \frac{m_{\rm Ar}}{m_{\rm Ge}} \cdot
    \frac{\mathcal{E}}{M_{\rm Ar}} \,.
\end{split}
\label{Eq:lifetime_limit_e}
\end{equation}
Here $N_{\rm e, Ge}=32$ and $N_{\rm e, Ar}=18$ are the numbers of electrons in Ge and Ar atoms. The LAr molar mass is $M_{\rm Ar} = 39.95\times10^{-3}$~kg/mol, with total mass $m_{\rm Ar}=3884.1$ kg. The total Ge mass $m_{\rm Ge}=38.78$ kg is computed as exposure-weighted averages of Phase~II and~II+ masses~\cite{GERDA:2020xhi}. Exposure $\mathcal{E}=105.5$ kg\,yr and efficiencies are taken from Table~\ref{tab:individual_exposure} and~\ref{tab:effi}, respectively. $M_{\rm tot}$ (kg/mol) is given in Sect.~\ref{sec:results:bDM}. \\

\begin{figure}[t!]
    \centering
    \includegraphics[width = 0.49\textwidth]{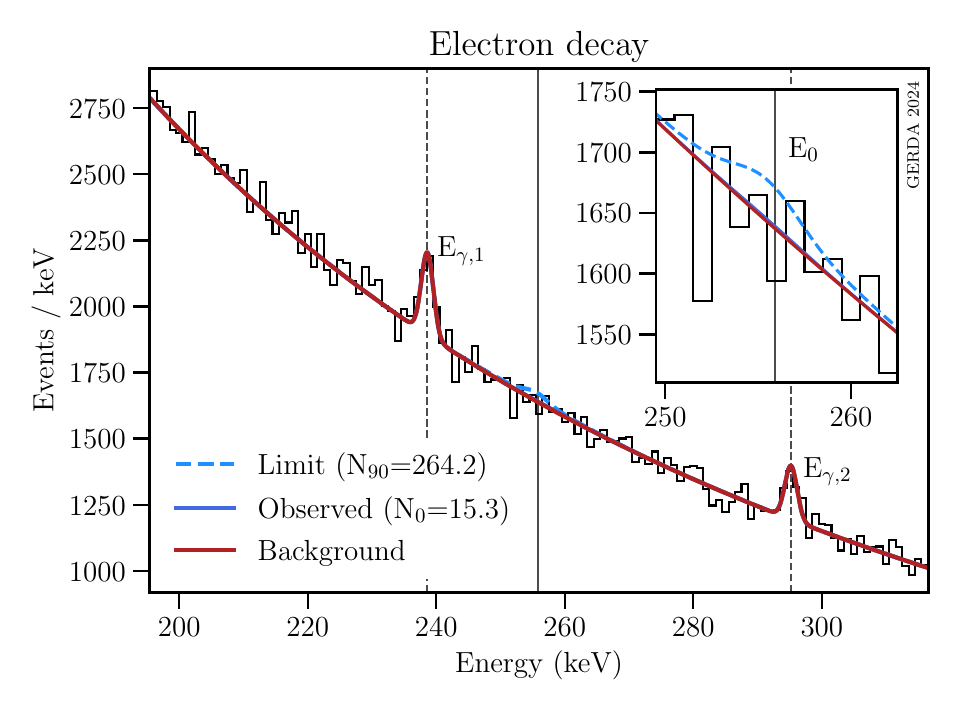} 
    \caption{
    Part of the M1 spectrum shown in Fig.~\ref{fig:M1data} with the Bayesian fit of the electron decay at $E_{\rm 0}=255.9$~keV (continuous line). The background fit includes two significant $\gamma$s (dashed lines) at $E_{\gamma,1}=$238.6~keV ($^{212}$Pb) and $E_{\gamma,2}=$295.2~keV ($^{214}$Pb), see Table~\ref{tab:detected_lines}
    } 
    \label{fig:fit_plot_electron}
\end{figure}

\noindent For the 255.9~keV Doppler broadened $\gamma$-line caused by a potential electron decay in Ge or Ar, no relevant deviation from the expected background was observed in the data. In the Bayesian fitting method, shown in Fig.~\ref{fig:fit_plot_electron}, the best-fit amplitude equals 15.3 counts with significance equal to $0.3\sigma$.
The obtained limit is $N_{\rm up,e}=264.2$ counts, and the median sensitivity is $N_{\rm s,e}=249.4$ counts. 
In the Frequentist procedure, a best-fit value of 3.8 counts is found, with vanishing significance. The evaluation of the upper limit yields $N_{\rm up,e}=263.1$ counts, with a sensitivity of $N_{\rm s,e}=259.2$ counts. 
The corresponding limits on the electron lifetime are listed in Table.~\ref{tab:lifetime_limits_recap}, and set into perspective in Table~\ref{tab:electron_exp_results}. The liquid-scintillator experiment Borexino set the currently tightest constraint. All other results were obtained with Ge detectors. Note that the validity of the statistical analysis conducted to obtain the numerical value of~\cite{Klapdor-Kleingrothaus:2007aoa} has been questioned 
in Refs.~\cite{Derbin:2007pz,PDG2022.upd2023}.
\begin{table}[t]
\caption{Selection of constraints on the electron lifetime $\tau_e$ at 90\%~CL
}
\begin{center}
\begin{tabular}{lccc}
\hline\noalign{\smallskip}
\textbf{Experiment} & \textbf{Nuclei} & \textbf{Decay} & $\boldsymbol{\tau_e}$\textbf{(yr)}  \\
\noalign{\smallskip}\hline\noalign{\smallskip}

   Borexino~\cite{BOREXINO:PRL115.231802} & C, H, N, O & $e^\text{-} \rightarrow \nu_{\rm e} \gamma$ & $6.6 \times 10^{28}$ \\

    HdM~\cite{Klapdor-Kleingrothaus:2007aoa}$^{(a)}$ & Ge & $e^\text{-} \rightarrow \nu_{\rm e} \gamma$ & $9.4 \times 10^{25}$\\

    \textsc{Majorana}~\cite{Majorana:2022mrm} & Ge & $e^\text{-}\rightarrow3\nu_{\rm e}$ &  $2.8 \times 10^{25}$ \\

    \textsc{Edelweiss}-III~\cite{PhysRevD.98.082004} & Ge & $e^\text{-}\rightarrow3\nu_{\rm e}$ & $1.2 \times 10^{24}$ \\

    \textcolor{blue}{\textsc{Gerda}} & \textcolor{blue}{Ge} & \textcolor{blue}{$e^\text{-} \rightarrow \nu_{\rm e} \gamma$} & \textcolor{blue}{$5.4 \times 10^{25}$} \\
\noalign{\smallskip}\hline
\multicolumn{3}{l}{
    \begin{minipage}{0.29\textwidth}
        $^{(a)}$ more likely overestimate~\cite{Derbin:2007pz,PDG2022.upd2023}
    \end{minipage}
}
\end{tabular}
\label{tab:electron_exp_results}
\end{center}
\end{table}

\section{Conclusions and outlook}
\label{sec:Conclusions}
In this paper, searches for full energy depositions caused by a coupling of bosonic DM with keV-scale masses with the atoms in the \textsc{Gerda} detectors are reported. 
No significant excess has been observed, hence constraints on the kinetic mixing of DPs as well as on the coupling of ALPs to electrons have been derived, in both Bayesian and Frequentist frameworks. Furthermore, the stability of the neutron and the proton inside $^{76}$Ge against inclusive decays with subsequent $\gamma$-only emission of the daughter isotope has been investigated by searching for a coincident signal induced by a $^{75}$Ge $\beta$ decay accompanied by the dominating $^{75}$As de-excitation $\gamma$-line of 264.60~keV. 
In addition, a Doppler broadened $\gamma$-line at 255.9~keV, which would be induced by the charge non-conserving decay of an electron into $\nu_{\rm e} \gamma$, has been analysed. 
None of the particle disappearance modes has been found, and constraints on the lifetimes of these particles have been derived in both statistical frameworks.\\
The limits for the search of DP and ALP DM pose the most stringent direct experimental results between roughly 140~keV and $2 m_{\rm e}$, 
except for masses in the 245-280 keV and 570-670 keV intervals where stronger constraints are set by COSINE-100~\cite{COSINE-100:2023dir}. However, for vector DM candidates, the indirect lifetime constraint based on the age of the Universe dominates significantly over the derived limits for masses above $\sim$500 keV.
In general, indirect galactic background searches for $3\gamma$ induced by DP decay are significantly more stringent~\cite{Redondo:2008ec}. In the energy range studied by \textsc{Gerda}, ALP DM models are mostly constrained by indirect, astrophysical measurements. Moreover, the ALP masses are further largely ruled out by the needed stability over the age of the Universe if one again assumes ALPs to compose the entire DM~\cite{PhysRevLett.128.221302}. The results for the ALP channel are shown as well, as more exotic, fine-tuned models have been suggested therein to omit the latter constraint. As a further remark, direct constraints on the absorption of ALPs have been reinterpreted to probe violations of \mbox{Poincaré} invariance~\cite{Gupta:2022qoq}. Hence, not only all combined results for ALPs and DPs, but also the individual absorption and 
the scattering channel constraints, are appended to this paper (see Fig.~\ref{fig:dark_alps_gerda_comparison}).
\\
\indent Regarding the determined lower lifetime limits on the inclusive nucleon decays in $^{76}$Ge, it is emphasised that, to our 
knowledge, these are the first constraints on these processes in $^{76}$Ge. However, the sensitivity of \textsc{Gerda} compared to the free nucleon 
decays or mode-dependent decays in any isotope is orders of magnitude below that  reached by large-scale experiments with light nuclei 
\cite{PhysRevLett.96.101802,Borexino:2003igu,PhysRevD.105.112012}.
The electron lifetime limit is among the strongest limits measured with semiconductor detectors, although 
the sensitivity does not reach that of large-scale organic scintillation experiments such as 
Bo\-re\-xino~\cite{BOREXINO:2017tdy}.
\\
The analyses presented here motivate further searches for these new physics channels with $\mathcal{O}(100\,\text{keV})$ energy depositions in semiconductor experiments. In particular, the future LEGEND-1000 experiment, aiming at the operation of more than one tonne of  Ge detectors enriched in $^{76}$Ge for ten years in underground-sourced LAr~\cite{LEGEND:2021bnm}, will improve these Ge-based constraints on bosonic DM interactions and lifetimes of electrons, neutrons, and protons.
The $^{39}$Ar concentration in under\-ground-sourced LAr is measured by the DarkSide collaboration to be reduced by a factor 1400~\cite{DarkSide:2018kuk}. Thus the sensitivity of LEGEND-1000 will be enhanced in the low-energy regime by more than an order of magnitude. 
Further improvements could be realised by deploying Ge detectors of natural isotopic composition (or depleted in $^{76}$Ge) in a setup similar to LEGEND-1000, to reduce the background induced by $2\nu\beta\beta$ decays.

\begin{acknowledgements}
The \textsc{Gerda} experiment is supported financially by the German Federal Ministry for Education and Research (BMBF), 
the German Research Foundation (DFG), 
the Italian Istituto Nazionale di Fisica Nucleare (INFN), 
the Max Planck Society (MPG), 
the Polish National Science Centre (NCN, Grant number UMO-2020/37/B/ST2/03905), 
the Polish Ministry of Science and Higher Education (MNiSW, Grant number DIR/WK/2018/08)
the Russian Foundation for Basic Research, and 
the Swiss National Science Foundation (SNF). 
This project has received funding and support from the European Union’s Horizon 2020 research and innovation programme under 
the Marie Sklodowska-Curie Grant Agreements No. 690575 and No. 674896. 
This work was supported by the Science and Technology Facilities Council, part of the U.K. Research and 
Innovation (Grant No. ST/T004169/1).
The institutions acknowledge also internal financial support. The \textsc{Gerda} collaboration thanks the directors and the staff of LNGS for their continuous strong support of the \textsc{Gerda} experiment.
\end{acknowledgements}

\begin{dataavailability}
This manuscript has associated data in a data repository. 
[Authors’ comment: The data shown in Figs. 3, 4 and 7 is available in ASCII format as Supplemental Material \cite{xyzdata}.]
\end{dataavailability}

\appendix
\section*{Appendix}
\renewcommand{\thesubsection}{\Alph{subsection}}
\renewcommand{\theequation}{\thesubsection.\arabic{equation}}
\subsection{Doppler broadened peak profile}
\label{sec:appendix_doppler_profile}
Using the virial theorem, i.e. $E_{\rm kin.} = - E_{\rm pot.}/2$, the Doppler broadened line shape can be analytically described as a sum of Gaussian contributions over all atomic shells weighted by their electron occupancy number $n_{\rm i}$,
\begin{equation}
    I(E) = \sum_{i=1}^{N_{\rm b}} I_{\rm i}(E) = \sum_{i=1}^{N_{\rm b}} \frac{n_{\rm i}}{\sqrt{2\pi}\sigma_{\rm i}}e^{-\frac{\left(E-E_{\rm t,i}\right)^2}{2\sigma_{\rm i}^2}}~,
\label{eq:doppler_line_shape}
\end{equation}
where $N_{\rm b}$ is the total number of atomic shells for a given atom~\cite{Klapdor-Kleingrothaus:2007aoa} and $E_{\rm t,i}$ is the total energy deposited in a detector after an electron decay (see Eqs.~\eqref{eq:el:inner_decay} and~\eqref{eq:el:outer_decay}). 
The line width for the \textit{i}-th atomic shell is
\begin{equation}
\sigma_{\rm i} = E_{\rm t,i} \cdot \sqrt{\frac{k_{\rm B} T_{\rm i}}{m_{\rm e}}} \approx 0.0442 \cdot E_{\rm t,i} \cdot \sqrt{E_{\rm b,i}} ~,
\label{eq:sigma_i_doppler}
\end{equation}
where $k_{\rm B}$ is Boltzmann’s constant and $T_{\rm i}$ is the absolute electron temperature, with energies $E_{\rm t,i}$ and $E_{\rm b,i}$ expressed in keV. Notice that the numerical pre-factor has been found upon recalculation, whereas \cite{Klapdor-Kleingrothaus:2007aoa} states a slightly larger value of 0.0447. The individual Ge and Ar atomic shell contributions as deduced from their respective electron binding energies are listed in Table~\ref{tab:lineshape_256_Ge_Ar}. 

\noindent Considering both Ge and Ar decays, the Doppler-broadened line shape is given as
\begin{equation}
    \label{eq:doppler_line_shape_GeAr}
    \begin{aligned}
    I(E) & \propto \,N_{\rm e,Ge} \cdot m_{\rm Ge} \sum_{i}^{N_{\rm b, Ge}}  I_{\rm i, det}(E) \cdot \epsilon_{\rm Ge, det} \\ 
    & + \,N_{\rm e,Ge} \cdot m_{\rm Ge} \sum_{i}^{N_{\rm b, Ge}}   I_{\rm i, mat} (E) \cdot \epsilon_{\rm Ge, mat} \\
    & + N_{\rm e,Ar} \cdot m_{\rm Ar}  \sum_{i}^{N_{\rm b, Ar}} I_{\rm i,Ar}(E) \cdot \epsilon_{\rm Ar} ~,
    \end{aligned}
\end{equation}
\noindent where $N_{\rm e,Ge}$ ($N_{\rm e,Ar}$) is the total number of available electrons in Ge (Ar) atoms, $m_{\rm Ge}$ ($m_{\rm Ar}$) is the total mass of the Ge array (Ar volume), and $\epsilon_{\rm Ge}$ ($\epsilon_{\rm Ar}$) is the detection efficiency in the Ge array of the outgoing photon following an electron decay originating within the Ge (Ar) volume (see Table~\ref{tab:effi} in Sect.~\ref{sec:data_efficiency}).
For germanium, sensitive detector contributions (\textit{det}) and contributions from surrounding detector material (\textit{mat}) are taken into account separately. 
\begin{table}[t]
\caption{Germanium and argon electron binding energies $E_{\rm b,i}$ for different atomic shells as taken from \cite{Carlson_bind_en} together with electron shell occupation
numbers $n_{\rm i}$. The corresponding FWHM contributions to the Doppler broadening of the electron decay signal are separately shown for the dominant contributions coming from Ge source detectors (K, L1-L3, M1-M5, N1-N2) and from the LAr (K, L1-L3, M1-M3). The FWHM value of each atomic shell was derived according to Eq.~\eqref{eq:sigma_i_doppler}}
\begin{center}
\begin{tabular}{cccccc}
\hline\noalign{\smallskip}
\textbf{Shell} & $\boldsymbol{n_{\rm i}}$ & \multicolumn{2}{c}{$\boldsymbol{E_{\rm b,i}}$ \textbf{(keV)}} & \multicolumn{2}{c}{\textbf{FWHM$\boldsymbol{_{\rm i}}$ (keV)}} \\
\cmidrule(lr){3-4} \cmidrule(lr){5-6} & & \textbf{Ge} & \textbf{Ar} & \textbf{Ge} & \textbf{Ar} \\
\noalign{\smallskip}\hline\noalign{\smallskip}
    K & 2 & 11.103 & 3.2059 & 90.6 & 47.4\\
    L1 & 2 & 1.4146 & 0.3263 & 31.7 & 15.2\\
    L2 & 2 & 1.2481 & 0.2506 & 29.8 & 13.3\\
    L3 & 4 & 1.217 & 0.2484 & 29.5 & 13.3\\
    M1 & 2 & 0.1801 & 0.0293 & 11.4 & 4.6\\
    M2 & 2 & 0.1249 & 0.0159 & 9.6 & 3.4\\
    M3 & 4 & 0.1208 & 0.0157 & 9.4 & 3.3\\
    M4 & 4 & 0.0298 & - & 4.8 & -\\
    M5 & 6 & 0.0292 & - & 4.8 & -\\
    N1 & 2 & 0.0143 & - & 3.2 & -\\
    N2 & 2 & 0.0079 & - & 2.4 & -\\
\noalign{\smallskip}\hline
\end{tabular}
\label{tab:lineshape_256_Ge_Ar}
\end{center}
\end{table}
\subsection{Empirical background model}
\label{sec:appendix_bkgr}
\begin{figure}[ht!]
    \centering
    \includegraphics[width = 0.45\textwidth]{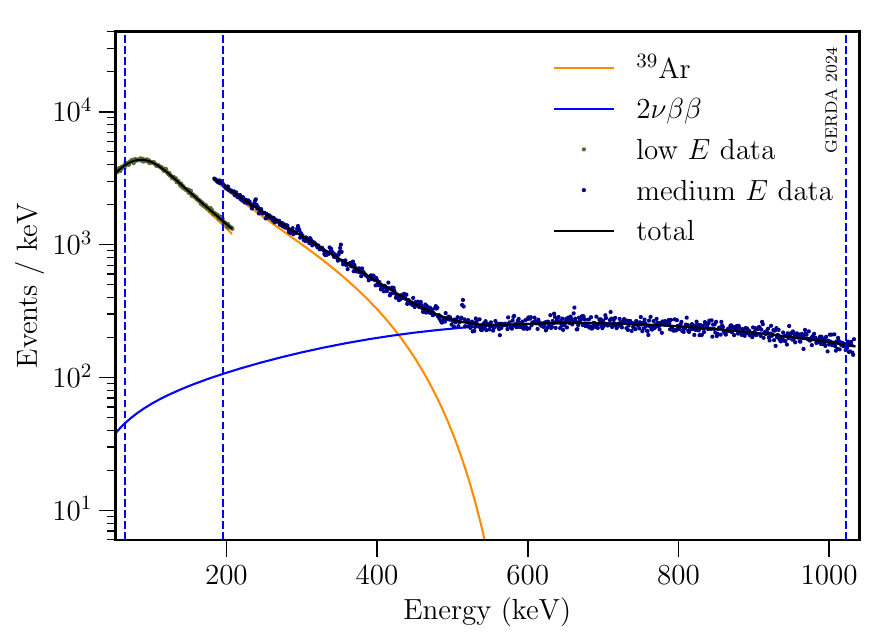}%
    \\
    \includegraphics[width = 0.45\textwidth]{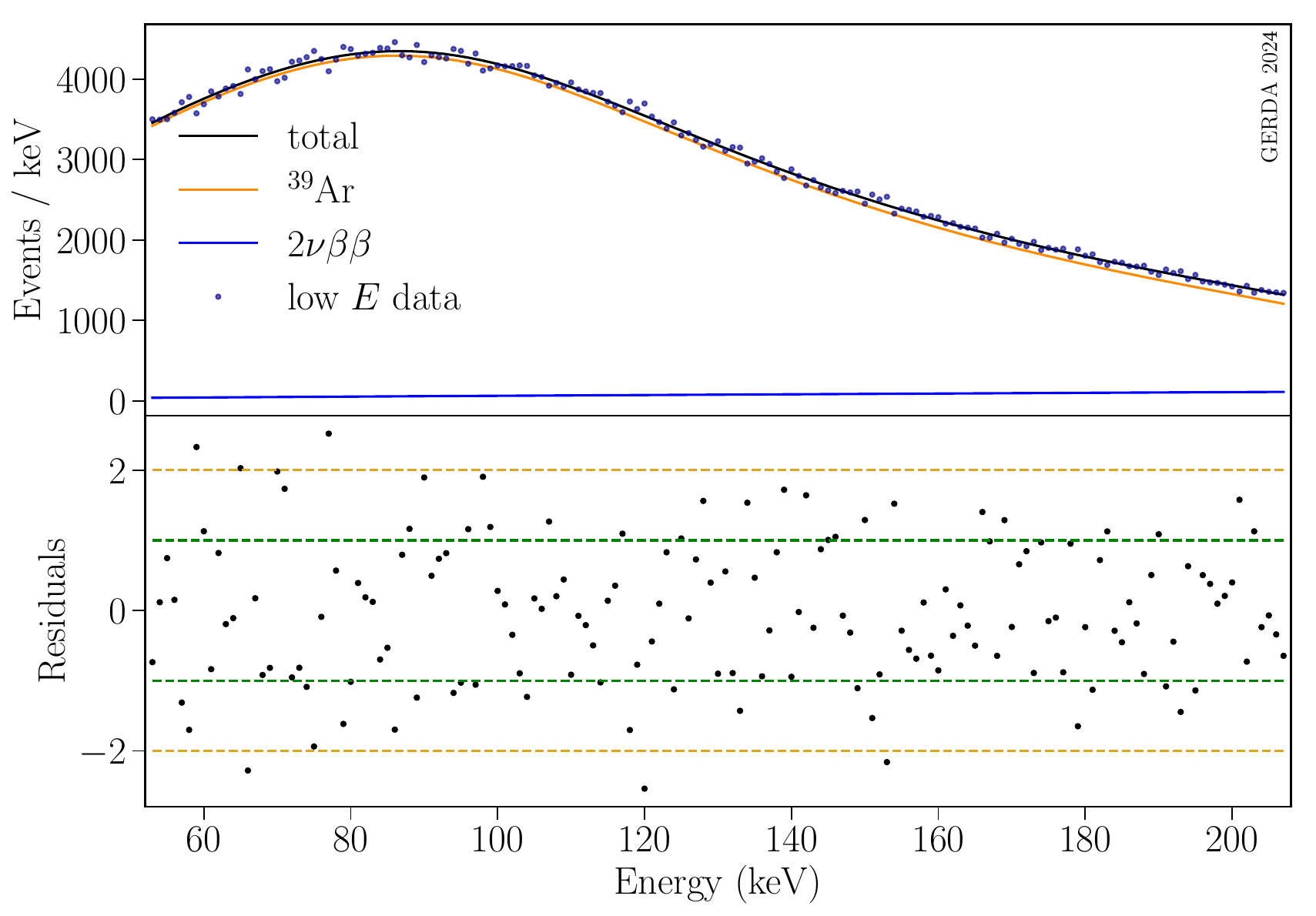}%
    \\
    \includegraphics[width = 0.45\textwidth]{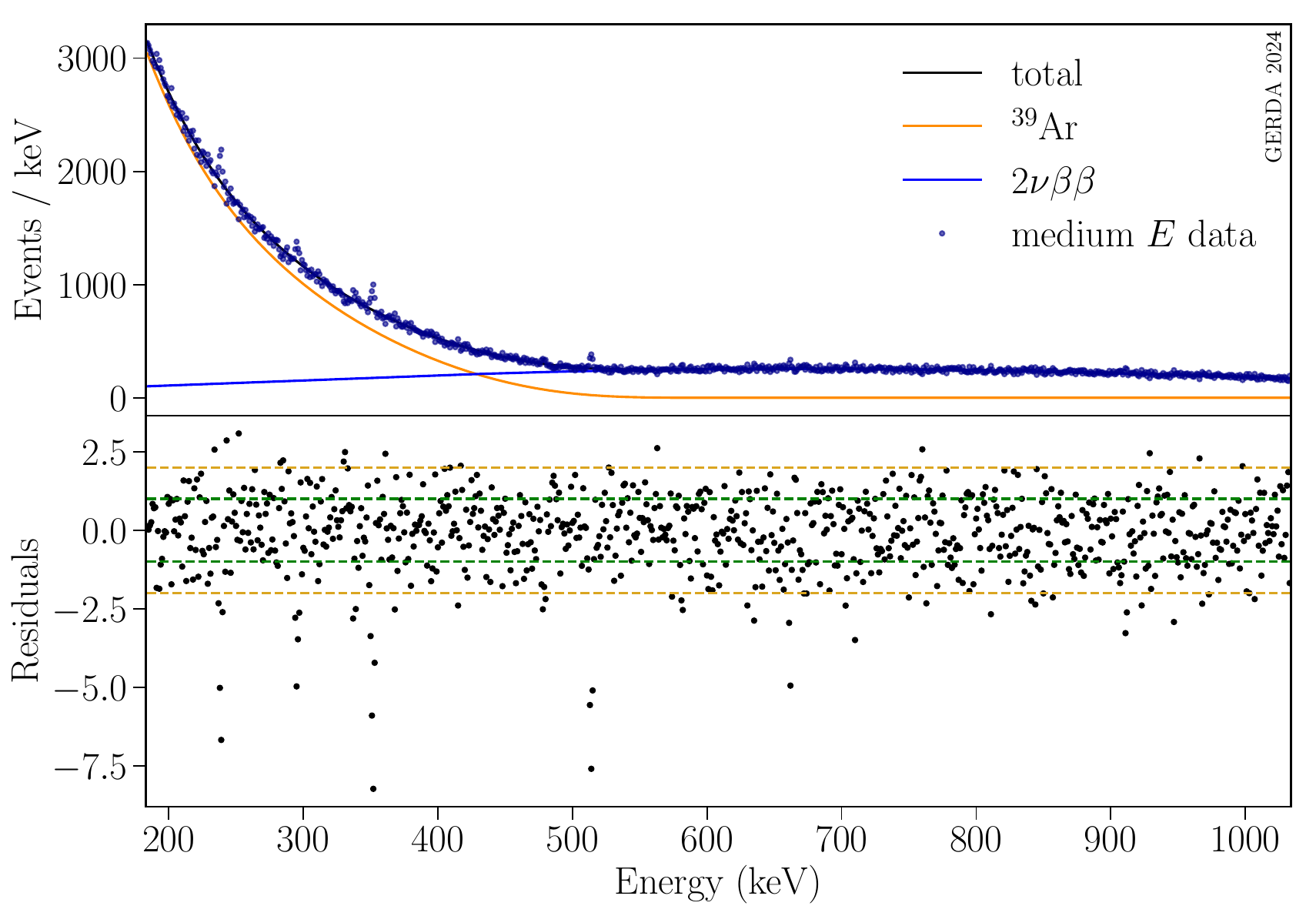}%
    \caption{Top: empirical background fit model. The fit was performed with a tenth order polynomial and a $\beta$-modified $\beta$ distribution. The vertical dashed, blue lines denote the lowest probed DM mass of 65~keV, the data set transition value of 195~keV, and 1021~keV as the highest potential integer DM mass below $2 m_{\rm e}$. Middle, bottom: plots of the data (blue dots) and the model (black line) in the two different energy ranges, i.e. 53-207~keV and 184-1033~keV, with the respective residuals shown below each panel. Residuals are defined as the difference between expected counts and observed counts, normalized by the square root of expected counts
    }
    \label{fig:bkgr_model}%
\end{figure}

The empirical background model, as well as its components (i.e. the $2\nu\beta\beta$ and the $^{39}$Ar decays), are shown in the top panel of Fig.~\ref{fig:bkgr_model}, together with the M1 data (see Sect.~\ref{subsec:Data selection}, Fig.~\ref{fig:M1data}) to which the model has been fit. A bin width of 1~keV was used, consistent with the analysis procedure presented in this paper. Figure~\ref{fig:bkgr_model} shows fits in two separate energy regions, i.e. 53 to 207~keV (middle) and 184 to 1033~keV (bottom), together with the corresponding residuals, defined as the difference between expected and observed counts over the square root of the expected counts. 
The two energy regions visible in the top panel were chosen such that to account for the 25 keV width of the fit window used in DM searches and to correctly handle the change in exposure around 195 keV due to the lowering of trigger thresholds in October 2017.

\noindent The empirical modified $\beta$ distribution modelling the $^{39}$Ar $\beta$ spectrum is 
based on Eq.~(5) of ref.~\cite{Awodutire:2021}, using a $\beta$ distribution as the baseline 
distribution. It was restricted to ten free parameters in this use-case: two shape parameters plus 
shift and scale parameters, for both $\beta$ components, one modification parameter, and one global 
amplitude parameter. For the empirical $2\nu\beta\beta$ distribution, modelled as a tenth-order 
polynomial vanishing at both 0 keV and the $Q_{\beta\beta}$ value, five parameters are kept free, 
analogously to the parametrization presented in \cite{Primakoff:1959chj}. The optimum parameters 
for both the $2\nu\beta\beta$ function and the $^{39}$Ar parametrization have been found via a 
combined histogram fit. Apart from the clear deviations at and around the observed $\gamma$-line 
positions as discussed in Sect.~\ref{subsec:Statistical frameworks}, the residuals largely 
fluctuate within the expected 1 and 2$\sigma$ ranges.

\noindent
The validity of the model was investigated using the reduced $\chi^2$/{\it dof} estimator where 
{\it dof} refers to the degrees of freedom. The fit yields $\chi^2$/{\it dof} $\approx$ 1.09 (1.51) 
for the low (high) energy data set from 53 to 207 (184 to 1033) keV. 
Including all identified $\gamma$-transitions in the high energy range (see Table 3) 
improves a posteriori the $\chi^2$/{\it dof} value to 1.06..
A further goodness-of-fit measure, the non-parametric \mbox{Kolmogorov}-\mbox{Smirnov} 
test~\cite{kolmogorov}, yields p-values p$_{KS}$ of 0.99 (0.16) for the low (high) energy data set 
(0.38 after including a posteriori the identified $\gamma$-lines). In summary, no significant 
deviations between 
the model and the data were found considering a posteriori all identified $\gamma$-transitions. 
We used the \mbox{Kolmogorov}-\mbox{Smirnov} test also to check the normality of the distribution of
the fit residuals. For the fit residuals of the low energy spectrum p$_{KS}$ equals 0.70. 
In the high energy range we find p$_{KS}$\,=\,0.005, or 0.46 when excluding identified $\gamma$-lines.
In conclusion, no significant deviation of the distribution of the residuals from normality was 
observed outside the locations of identified $\gamma$-lines.

\subsection{Statistical frameworks}
\label{sec:appendix_statistical_frameworks}
In this section, the applied statistical methods are described in detail.
\paragraph{Bayesian method}
To identify a potential excess at any probed energy value, a binned Bayesian fit of the signal peak above the background was performed in the respective signal window. Poisson fluctuations were assumed for bin contents. The Markov-Chain-Monte-Carlo algorithm was applied via the Bayesian Analysis Toolkit (BAT) software~\cite{Caldwell:2008fw}. A uniform prior was chosen to constrain the signal amplitude to the physically allowed positive range. The posterior signal distribution was then marginalised via eight Markov chains of $10^6$ iterations each. The significance of signal strengths having a marginalized posterior distribution incompatible with zero counts was estimated via the global mode divided by the upper and the lower 68\% quantiles of the posterior distribution, $\sigma=\frac{U_{68}-L_{68}}{2}$.
Defining the significance in this manner, the maximally visible excess at 710~keV (see Sect.~\ref{subsec:Statistical frameworks}), which cannot be attributed to an expected $\gamma$-line, has a significance of $2.9\sigma$. 

\paragraph{Frequentist method}
For the fitting procedure in the Frequentist statistical framework, 
the local significance was estimated for each of the probed DM candidate masses assuming the asymptotic $\frac{1}{2}\chi^2(1)$ distribution, cf. \cite{Cowan:2010js}, where 1 denotes the degrees of freedom. The unexpected excess at 710~keV (see Sect.~\ref{subsec:Statistical frameworks}) has a local significance of $3.3\sigma$. 
Given the large number of searches, this estimate needs to be corrected for the look-elsewhere effect. The compensation of this effect can be approximated by applying a \mbox{Bonferroni} correction~\cite{Bonferroni1936teoria}, meaning a rescaling of the local p-values by the number of trials. A less conservative option is the method of data-driven self-calibration~\cite{Bayer:2021lhk}. The global significance estimation in this method is based on peaks artificially induced into the data. Upon both \mbox{Bonferroni} correction and self-calibration, the observed $3.3\sigma$ peak corresponds to a global significance $\leq\!1\sigma$, and might be interpreted as a noise fluctuation. Alternatively, this peak might be of physical origin, i.e. caused by the presence of an unexpected isotope in or near the Ge detectors.\\ 
The determined limits were obtained with the profile likelihood ratio method~\cite{Rolke:2004mj}, partially via the MINUIT2 algorithm~\cite{James:1975dr}. The test statistics $\tilde{t}$ of \cite{Cowan:2010js} was applied to constrain the physical signal strength to positive values, again relying on the asymptotic (non-central) $\chi^2(1)$ distributions. The median exclusion sensitivity and the non-centrality parameter were estimated from the Asimov data set, as motivated in \cite{Cowan:2010js} as well.
\subsection{Direct dark matter absorption vs dark Compton scattering}
\label{sec:appendix_bDM_comparison}
\begin{figure}[t]
  \includegraphics[width = 0.49\textwidth]{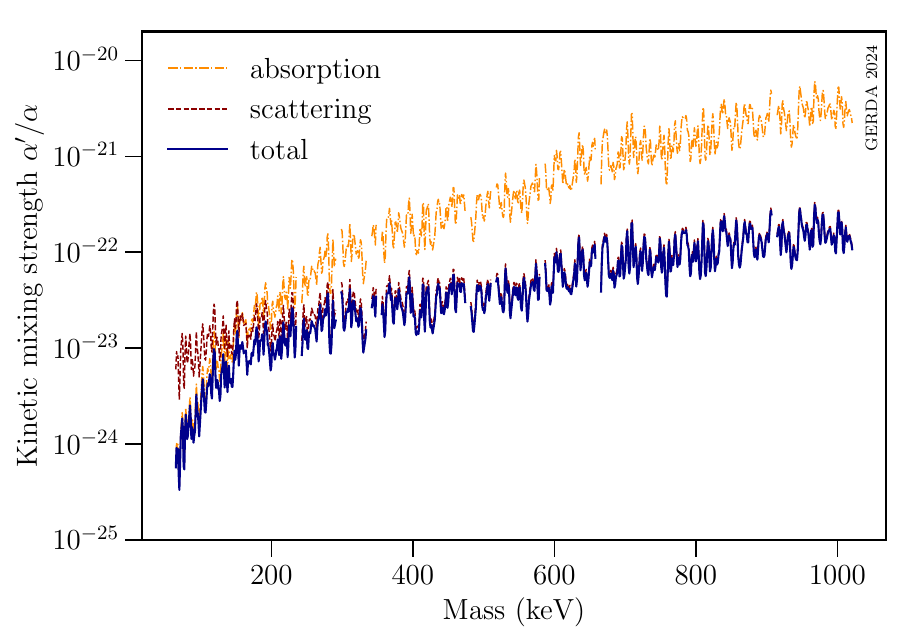}
\caption{Comparison of Bayesian limits at 90\%~CI for the dimensionless coupling constant of DPs to electrons, plotted as a function of the respective DM mass when 
evaluated by considering photoelectric-like absorption only (gold), Compton scattering only (red), and both interactions (blue). Regions around identified 
$\gamma$ lines (see Sect.~\ref{subsec:Statistical frameworks}, Table~\ref{tab:detected_lines}) were omitted 
}%
    \label{fig:dark_alps_gerda_comparison}%
\end{figure}

Fig.~\ref{fig:dark_alps_gerda_comparison} compares the effect of direct dark matter absorption and dark Compton scattering on the Bayesian limit for the kinetic mixing coupling of DPs to electrons. 
Including the dark Compton scattering interaction induces a strong sensitivity improvement compared to the previous results~\cite{GERDA:2020emj} at higher energies. The same conclusions hold for the limits on the ALP-electron coupling strengths (not shown). Table~\ref{tab:bDM_example_results} shows selected results on the kinetic mixing strength of DPs and the coupling of ALPs to electrons taking both 
direct dark matter absorption and dark Compton scattering into account.

\begin{table*}[ht!]
\caption{Bosonic DM upper limits (${L}$) and sensitivities (${S}$) at 90\% CI/CL on the kinetic mixing strength of DPs (${\alpha^{'}/}{\alpha}$) and on the coupling of ALPs to electrons ($g_{\rm ae}$) at indicated masses as determined in the Bayesian and Frequentist frameworks. The photoelectric-like absorption process as well as the dark Compton scattering were included in the DM interaction rate with Ge material when deriving the coupling values. For each mass, the observed best-fit value (${obs.}$) is shown together with its significance ($sig.$).
For non-positive ${obs.}$ values, the significance is null and not displayed. 
The extracted upper limits at 90\% CI/CL and the median sensitivity for the signal strength are indicated with $N_{\rm up}$ and $N_{\rm s}$, respectively.
Upper limits derived for all masses between 65~keV and $2m_{\rm e}$ are shown in Fig.~\ref{fig:dark_alps}. Sensitivities for the entire mass range are shown in Fig.~\ref{fig:dark_alps_gerda_comparison_stats_sensitivity}}
\begin{center}
\label{tab:bDM_example_results} 
\begin{tabular}{c c ccc cc cc}
\hline\noalign{\smallskip}
\textbf{Mass} & \textbf{Framework} & \multicolumn{3}{c}{\textbf{Signal counts}} & \multicolumn{2}{c}{$\boldsymbol{{\alpha^{'}/}{\alpha}}$ \textbf{(DPs)}} & \multicolumn{2}{c}{$\boldsymbol{g_{\rm ae}}$ \textbf{(ALPs)}}\\  
\cmidrule(lr){3-5} \cmidrule(lr){6-7} \cmidrule(lr){8-9} \textbf{(keV)} && \textbf{$\boldsymbol{obs.\,(sig.)}$} & \textbf{$\boldsymbol{N_{\rm up}}$} & 
\textbf{$\boldsymbol{N_{\rm s}}$} & \textbf{$\boldsymbol{L}$} & \textbf{$\boldsymbol{S}$} & \textbf{$\boldsymbol{L}$} & \textbf{$\boldsymbol{S}$} \\
\noalign{\smallskip}\hline\noalign{\smallskip}
    65 & Bayesian & 22.2 (0.5$\sigma$) & 189.7 & 173.2 & $5.7 \times 10^{-25}$ & $5.2 \times 10^{-25}$ & $2.0 \times 10^{-12}$ & $1.9 \times 10^{-12}$ \\
    196 & & 23.8 (0.9$\sigma$) & 171.9 & 161.4 & $1.1 \times 10^{-23}$ & $9.9 \times 10^{-24}$ & $2.8 \times 10^{-12}$ & $2.7 \times 10^{-12}$ \\
    1021 & & 0.0 & 34.4 & 46.0 & $1.1 \times 10^{-22}$ & $1.4 \times 10^{-22}$ & $3.3 \times 10^{-12}$ & $3.8 \times 10^{-12}$ \\ \hline
    
    65 & Frequentist & -89.4  & 99.6 & 177.2 & $3.0 \times 10^{-25}$ & $5.4 \times 10^{-25}$ & $1.5 \times 10^{-12}$ & $2.0 \times 10^{-12}$ \\ 
    196 & & ~50.5 (0.5$\sigma$) & 210.9 & 159.7 & $1.3 \times 10^{-23}$ & $9.9 \times 10^{-24}$ & $3.1 \times 10^{-12}$ & $2.7 \times 10^{-12}$ \\
    1021 & & -15.9 & 31.0 & 45.8 & $1.0 \times 10^{-22}$ & $1.4 \times 10^{-22}$ & $3.1 \times 10^{-12}$ & $3.8 \times 10^{-12}$ \\
\noalign{\smallskip}\hline
\end{tabular}
\end{center}
\end{table*}

\subsection{Comparison of bosonic dark matter sensitivities}
\label{sec:appendix_bDM_comparison_stats}
\begin{figure}[h!]
  \includegraphics[width = 0.49\textwidth]{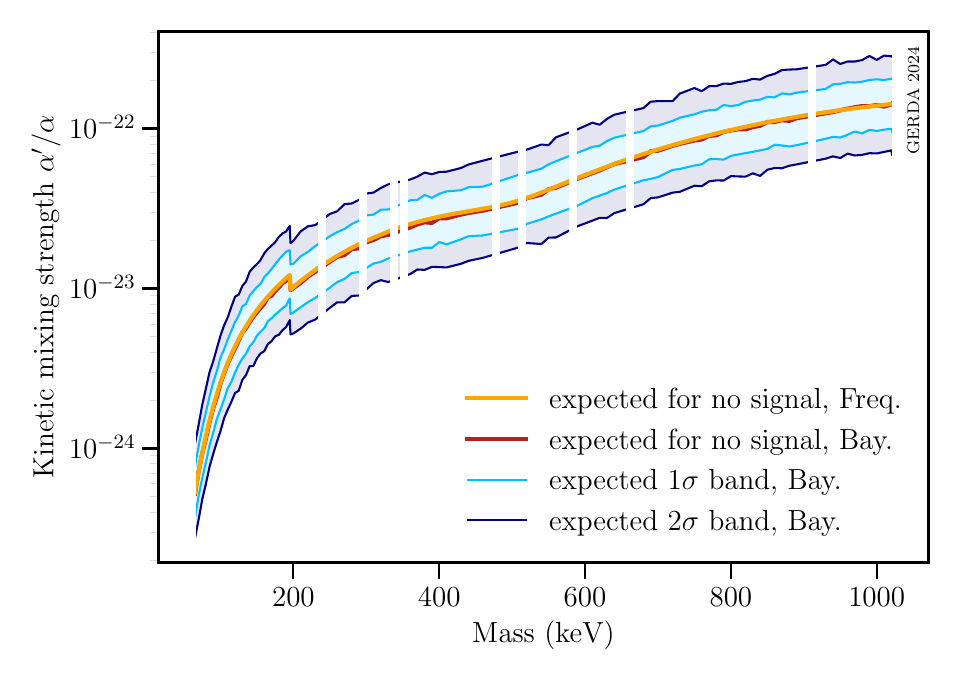}
\caption{Comparison of Bayesian (red) and Frequentist median sensitivities (gold) for the dimensionless coupling constant of DPs, 
  plotted as a function of the respective DM masses. Couplings here are evaluated considering photoelectric-like absorption and 
  Compton scattering processes. The indicated blue bands correspond to the 1 and 2$\sigma$ range for the Bayesian limits, 
  respectively. Regions around identified $\gamma$ lines (see Sect.~\ref{subsec:Statistical frameworks}, 
  Table~\ref{tab:detected_lines}) were omitted} %
\label{fig:dark_alps_gerda_comparison_stats_sensitivity} %
\end{figure}

The Bayesian (Frequentist) median sensitivities assuming no signal are plotted for the kinetic mixing coupling of DPs to electrons in Fig.~\ref{fig:dark_alps_gerda_comparison_stats_sensitivity}, together with the expected 1 and 2$\sigma$ fluctuation bands for the Bayesian limits, as determined from a set of $10^3$ MC simulations sampled individually at each inspected integer mass value. 
Here, both the photoelectric-like absorption and Compton scattering processes are taken into account when extracting the coupling values. 
The Frequentist sensitivities were extracted directly from the Asimov data sets (see Sect.~\ref{sec:appendix_statistical_frameworks}).
The drop visible around 196 keV is related to the difference in exposure between the energy intervals of 65-195 keV (45.5 kg\,yr) and 196-1021 keV (60.0 kg\,yr). 
Upper limits shown in Fig.~\ref{fig:dark_alps_gerda_comparison} lie well within the expectation bands. 
The same behaviour is found for ALP-electron coupling strengths (here not shown). 



\end{document}